\documentclass[hidelinks]{article}
\usepackage[utf8]{inputenc}
\usepackage[left=3cm,right=3cm,top=3cm,bottom=3cm]{geometry}
\usepackage{pdfpages}
\usepackage{hyperref}
\usepackage{listings}
\usepackage{xcolor}
\usepackage{setspace}
\usepackage{tocloft}
\usepackage{amsmath}
\usepackage{chngcntr}
\usepackage[toc,page]{appendix}
\usepackage[T1]{fontenc}
\usepackage[nottoc]{tocbibind}
\usepackage[compact]{titlesec}
\usepackage[utf8]{inputenc}
\usepackage{natbib}
\usepackage{amsmath}
\usepackage{amssymb}
\usepackage{amsfonts}

\usepackage{multirow}
\usepackage{setspace}
\usepackage{array}
\usepackage{float}

\newcolumntype{L}[1]{>{\raggedright\let\newline\\\arraybackslash\hspace{0pt}}m{#1}}
\newcolumntype{C}[1]{>{\centering\let\newline\\\arraybackslash\hspace{0pt}}m{#1}}
\newcolumntype{R}[1]{>{\raggedleft\let\newline\\\arraybackslash\hspace{0pt}}m{#1}}

\usepackage{subfig}
\usepackage{bm}
\usepackage{multirow}
\usepackage{tikz}
\usetikzlibrary{shapes,arrows}
\usepackage{blkarray}
\usepackage[affil-it]{authblk}
\usepackage{listings}
\usepackage{siunitx}
\usepackage{lineno}

\usepackage{natbib}
\usepackage{color}
\providecommand{\keywords}[1]{\textbf{\textit{Keywords ---}} #1}

\title{A Bayesian aoristic logistic regression to model spatio-temporal crime risk under the presence of interval-censored event times}
\date{}
\author[1*]{Álvaro Briz-Redón}
\affil[1]{Department of Statistics and Operations Research, University of Valencia, Spain}
\affil[*]{\textsf{alvaro.briz@uv.es}}

\begin{document}

\maketitle

\begin{abstract}

From a statistical point of view, crime data present certain peculiarities that have led to a growing interest in their analysis. In particular, a characteristic that some property crimes frequently present is the existence of uncertainty about their exact location in time, being usual to only have a time window that delimits the occurrence of the event. There are different methods to deal with this type of interval-censored observation, most of them based on event time imputation. Another alternative is to carry out an aoristic analysis, which is based on assigning the same weight to each time unit included in the interval that limits the uncertainty about the event. However, this method has its limitations. In this paper, we present a spatio-temporal model based on the logistic regression that allows the analysis of crime data with temporal uncertainty, following the spirit of the aoristic method. The model is developed from a Bayesian perspective, which allows accommodating the temporal uncertainty of the observations. The model is applied to a dataset of residential burglaries recorded in Valencia, Spain. The results provided by this model are compared with those corresponding to the complete cases model, which discards temporally-uncertain events.

\end{abstract}


\newpage

\section{Introduction}

The use of advanced statistical techniques for crime analysis has experienced significant growth in the last decade. In particular, it is of special interest to analyze the distribution of crime in space and time, and also to study the presence of space-time interaction. For this reason, different modeling approaches have been developed and adapted to find environmental factors that are associated with a higher risk of crime, as well as to predict crimes in the short-term and the mid-term. Among others, different versions of self-exciting models \citep{mohler2011self,zhuang2019semiparametric}, spatial models with a non-linear structure \citep{briz2022identifying}, and Bayesian univariate/multivariate spatio-temporal models \citep{chung2019crime,law2014bayesian,li2014space,quick2018crime} have been proposed. 

Crime data often presents singular characteristics that can complicate the analysis, which, in turn, leads to the development of new methodologies. For instance, it is well acknowledged that crime figures are usually underestimated, or even biased \citep{buil2021measuring,buil2021accuracy}. Besides, in the context of spatial, temporal, or spatio-temporal crime datasets, both the spatial and the temporal accuracy are often a matter of concern. Problems with spatial accuracy usually refer to the impossibility of identifying the spatial unit over which the event has occurred (considering, for example, an administrative division of the study area). In particular, if the spatial location of the event is available in the form of textual information (representing the human-readable address of the location), it is common to have geocoding errors or events that fail to be geocoded. This type of problem has been discussed previously, with the aim of establishing minimally acceptable geocoding rates \citep{andresen2020minimum,briz2020reestimating,ratcliffe2004geocoding}. Regarding the issues related to temporal accuracy, it is also usual that for some crime events we do not observe their temporal location with the desired accuracy (minute, hour, date, etc.). Indeed, this situation takes place in most events if we are dealing with certain types of crime such as property theft \citep{ashby2013comparison}. In this case, what we have is a lower and an upper bound of the temporal location of the event, that is, a temporal interval or window for each of the events. This kind of temporal observation is usually referred to as interval-censored.

The existence of interval-censored event times is a well-known issue in the field of quantitative criminology. Even though it may not have received enough attention, there are different strategies to deal with interval-censored event times in criminal records. The simplest approach is to choose an appropriate time unit that eliminates temporal uncertainty. For example, if the existing uncertainty is at the day level, we can operate at the week level. Although in some cases this approach allows us to avoid the temporal uncertainty of most (or all) observations, setting the temporal resolution of the analysis based on the uncertainty surrounding the observations is not desirable. 

Another way to deal with uncertainty is to perform the imputation of event times, which is possibly the most common strategy and can be performed in several ways. One possibility is to assign to these events the temporal unit (hour, date, etc.) that lies just at the midpoint of the uncertainty time window available. Considering the initial or the final temporal location of the time window is another option, but the midpoint approach has less error (and bias) associated \citep{ashby2013comparison}. However, these deterministic methods are outperformed by a non-deterministic one called the aoristic method, as shown by \cite{ashby2013comparison}. 

The aoristic method, which has been proposed mainly by criminologist J. Ratcliffe \citep{ratcliffe1998aoristic,ratcliffe2000aoristic,ratcliffe2002aoristic}, is based on assigning the same weight to each time unit included in the interval that delimits the uncertainty about the crime event. Hence, when one uses the aoristic approach, temporally-uncertain events do not receive a single imputation value, but a probability score for each of the temporal units within which the event is located. Specifically, all the temporal units receive the same score or weight, so that they add up to 1. For this reason, the aoristic method does not entirely correspond to an imputation method. Nevertheless, by following the aoristic procedure, one can, for instance, deduce the temporal distribution of a set of crimes by adding both the number of certain events and the fractions of uncertain events corresponding to each date within the period. Even though the aoristic approach allows carrying out some exploratory analyses of crime datasets including interval-censored temporal observations, this may not be sufficient depending on the purpose of the analysis. 

Finally, model-based approaches can also be followed to deal with interval-censored observations. For instance, the von Mises distribution or Dirichlet processes, which are typically used for the analysis of circular data, have been recently proposed for the analysis of aoristic data \citep{mulder2019bayesian}. Furthermore, one way to deal with the temporal uncertainty of the observations is to include all of them in a single model and assume a certain probability distribution on the exact temporal location of these observations. This leads, in general, to models that handle missing data under the Bayesian framework. Indeed, \cite{reich2015partially} adopted a Bayesian modeling framework for clustering criminal events which, among other features, allows for dealing with interval-censored event times. In this case, the authors treated interval-censored event times as latent variables with full conditional following a truncated normal distribution with mean equal to the cluster mean (that is, the average temporal location of the events belonging to the same cluster). 

In this paper, the latter approach is followed to analyze a burglary dataset recorded in Valencia, Spain, which, as is common with residential burglaries, includes a large proportion of temporally-uncertain events. Specifically, a spatio-temporal logistic regression is proposed to model burglary risk in space and time. The model is estimated within a Bayesian framework, allowing the inclusion of temporally-uncertain events. The aoristic approach is imitated when introducing this uncertainty into the model. Therefore, the objective of the paper is twofold. First, to describe a modeling framework to estimate burglary risk in space and time, while accounting for events with temporal uncertainty. Second, to highlight the suitability of including this kind of crime event in the analysis to get more reliable parameter estimates and attempt to recuperate the underlying spatio-temporal distribution of crime. 

The paper is structured as follows. Section \ref{Data} contains a description of the data used for the analysis, emphasizing the presence of temporal uncertainty in the events. Section \ref{Methodology} describes the modeling framework proposed for the estimation of burglary risk in space and time, under the presence of interval-censored temporal observations. Then, Section \ref{Results} shows the main results derived from the analysis. Finally, Section \ref{Discussion} includes a discussion and some concluding remarks.

\section{Data}
\label{Data}

\subsection{Study settings}

The case study has been conducted in the city of Valencia, the third most populated city in Spain, with a population size of around 800,000 inhabitants. Specifically, the urban core of the city has been considered for the analysis, excluding some peripheral districts that only represent 5\% of the population. Besides, to investigate the spatial distribution of burglaries across the city, the boroughs of Valencia have been considered for analysis. There are 70 boroughs in the study area delimited for the research.

\subsection{Burglary data}

A dataset provided by the Spanish National Police containing information about 2626 geocoded burglaries recorded in the city of Valencia from 1 January 2016 to 31 December 2017 has been used for the analysis. This dataset has already been analyzed by \cite{briz2020adjusting} to study the near-repeat phenomenon. In this dataset, the geographical coordinates are available for each of the events, allowing the analysis to be conducted at any desired spatial scale. In contrast, the temporal location of some of the burglary events has a certain degree of uncertainty. Particularly, the exact date of occurrence of the burglary is known for only the $60.9\%$ of the events. Further comments on the temporal uncertainty of the data are provided in the following Subsection.

\subsection{Event time uncertainty}

One important feature of the dataset under study is the presence of interval-censored event times. Specifically, for each burglary, there is a \textit{from date} and a \textit{to date} variable that allow delimiting the temporal location of the event, based on the information available about the burglary (the \textit{from date} represents the last date on the calendar on which the owners can be sure that the home has not yet been burgled, whereas the "to date" is the date on which the owners, the Police, or any citizen has ascertained that the burglary has been committed). Although the original variables including dates are in a YYYY-MM-DD format, they have been transformed into numeric values to ease their use, assigning a value of 1 to 1 January 2016, which represents the start of the study period. In the remainder of the paper, the dates would be considered as numeric values, unless otherwise stated.

Thus, while there is no spatial uncertainty in the data since the coordinates of each dwelling that has been burgled during the period under study are available (of course, for the burglaries that have been notified to the Police), event time uncertainty cannot be overlooked. In a previous study by \cite{briz2020adjusting}, the temporal uncertainty of some observations was resolved through the midpoint date imputation method. A preliminary analysis conducted in the context of that study, focused on the near-repeat phenomenon, allowed concluding that the imputation method (midpoint date or aoristic date) did not have a strong impact on the results. Therefore, the midpoint date method was preferred over the aoristic given its computational convenience. 

In the present paper, the aim is to deal with interval-censored events explicitly, without direct imputation of missing event dates, as will be shown in the subsequent Section. 


\section{Methodology}
\label{Methodology}

\subsection{Case-control study design}

In order to follow a logistic modeling framework (which will be described in the next Subsection), a binary response variable indicating the presence or absence of a burglary event is needed. Therefore, the spatio-temporal locations (with event time uncertainty) of the burglaries available, denoted by $\{\boldsymbol{x}_i,t^{from}_{i},t^{event}_{i},t^{to}_{i}\}_{i=1}^{2626}$, are treated as the cases/events, where $\boldsymbol{x}_i$ are the geographical coordinates corresponding to event $i$, and $t^{from}_{i}$, $t^{event}_{i}$, and $t^{to}_{i}$ represent, respectively, the \textit{from date}, the date at which the event actually occurred, and the \textit{to date}. For events with no temporal uncertainty, it holds $t^{from}=t^{event}=t^{to}$. For temporally-uncertain events, we have $t^{from}<t^{to}$ and a missing value for $t^{event}$. In other words, we only know that $t^{event}\in[t^{from},t^{to}]$.

\begin{figure}[htbp]
 \centering
 \includegraphics[width=8cm,angle=-90]{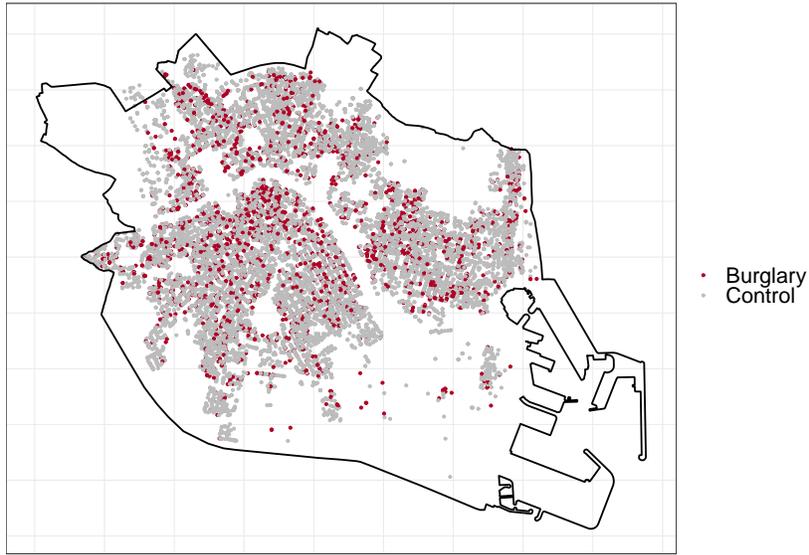}
\caption{Location within the study area of the events (burglaries) and controls considered for the analysis}
\label{fig:case_control}
\end{figure}

To generate the controls, the point pattern formed by the locations of all the dwellings in the study area is taken into account. Specifically, this pattern consists of 28682 locations within the study area, which include 382539 dwelling units (this corresponds to the dwellings registered in Valencia in 2016). A total of 13130 control locations were sampled with replacement, setting the probability of selection to be proportional to the number of dwellings in the location. Hence, the number of controls was chosen to be five times the number of cases, so a 5:1 ratio of controls to cases was used. The literature suggests that a 4:1 case-control ratio is generally sufficient to carry out a case-control study design \citep{gail1976many,hong2012sample}. The choice of this ratio will affect probability estimates derived from the logistic regression model, so this should be taken into account when performing binary (event/no event) predictions that depend on a cutoff probability. Each of these control locations was assigned a random date from 1 January 2016 to 31 December 2017. No temporal uncertainty has been assumed for the control data, so $t^{from}=t^{event}=t^{to}$ for all these space-time locations. Figure \ref{fig:case_control} displays the spatial locations of the cases and controls considered for the analysis over the study area.

\subsection{Logistic regression model}

If $\{\boldsymbol{x}_i,t^{from}_{i},t^{event}_{i},t^{to}_{i}\}_{i=1}^{15756}$ denotes the complete set of spatio-temporal locations considered for the analysis, let $y_i$ be a binary variable indicating if each location represents a case (burglary) or a control. As usual, we set $y_i=1$ if $i$ is an event, and 0 otherwise. In order to model the risk of burglary for each spatio-temporal location within the study window, a logistic regression modeling framework is a natural choice. Under the logistic model, the occurrence of a burglary event at spatio-temporal location $i$ is described through a Bernoulli random variable $Y_i\sim Ber(\pi_i)$, where $\pi_i$ represents the probability of occurrence of the burglary. In this paper, we attempt to explain this parameter in terms of several fixed and random (spatial and temporal) effects, leading to the following expression:
\begin{equation}
\mathrm{logit}(\pi_i)=\log\bigg(\frac{\pi_i}{1-\pi_i}\bigg)=\alpha+\beta_{DoW(i)}+\delta_{w(i)}+\varepsilon_{w(i)}+u_{b(i)}+v_{b(i)}
\label{eq:logistic_model}
\end{equation}
where each term is defined as follows. First, $\alpha$ is the global parameter of the logistic regression model. Second, $\beta_{DoW(i)}$ represents the effect of the day of the week corresponding to event $i$ ($DoW(i)$) on $\mathrm{logit}(\pi_i)$. Particularly, Monday is taken as the reference level of this variable, so six $\beta_{DoW}$ parameters are estimated, one for each of the remaining days of the week. A vague Gaussian prior, $N(0,1000)$, is assigned to $\alpha$ and the $\beta_{DoW}$'s. The rest of the parameters involved in the model represent spatial and temporal random effects, which are outlined in the following lines.

The temporally-structured week effect, $\delta_{w}$ ($w=1,...,104$), is specified through a second-order random walk
$\delta_{w}|\delta_{w-1},\delta_{w-2} \sim N(2\delta_{w-1}+\delta_{w-2},\sigma^2_{\delta})$, where $\sigma^2_{\delta}$ is the variance component, whereas an independent and identically distributed Gaussian prior is chosen for the temporally-unstructured week effect, $\varepsilon_{w} \sim N(0,\sigma^2_{\varepsilon})$ ($w=1,...,104$). The variance components, $\sigma^2_{\delta}$ and $\sigma^2_{\varepsilon}$, are assigned a Gamma-distributed prior, $Ga(1,0.5)$.

Regarding the spatial random effects, the Besag–York–Molliè (BYM) model has been employed \citep{besag1991bayesian}, under which the conditional distribution of the spatially-structured effect on a borough $b$, $u_{b}$ ($b=1,...,70$), is
$$u_{b}|u_{\Tilde{b} \neq b} \sim N\bigg(\sum_{\Tilde{b} \neq b=1}^{n} w_{b\Tilde{b}}u_{\Tilde{b}},\frac{\sigma^2_{u}}{N_{b}}\bigg)$$
where $N_{b}$ is the number of neighbors for area $b$, $w_{b\Tilde{b}}$ is the $(b,\Tilde{b})$ element of the row-normalized neighborhood matrix ($w_{b\Tilde{b}}=1/N_b$ if boroughs $b$ and $\Tilde{b}$ share a geographical boundary, and 0 otherwise), and $\sigma^2_{u}$ represents the variance of this random effect. The spatially-unstructured effect over the areas, denoted by $v_{b}$ ($b=1,...,70$), follows a Gaussian distribution, $v_{b} \sim N(0,\sigma^2_{v})$, where $\sigma^2_{v}$ is the variance of the effect. It was assumed a Gamma-distributed prior, $Ga(1,0.01)$, for both $\sigma^2_{u}$ and $\sigma^2_{v}$.

At this point, it should be noted that the choice of the structure formed by the boroughs of Valencia to measure the spatial variation of burglary risk is to some extent arbitrary. In this case, the choice of an administrative unit is convenient because it would allow practitioners to design surveillance strategies in a simple way. Moreover, from a computational point of view, it is also advantageous, due to the not too large number (70) of spatial units contained. In any case, the type of modeling framework proposed could be adapted to other types of partitions, regular or irregular, more or less fine, of the study area under consideration.



\subsection{Dealing with the temporal uncertainty}

The logistic regression model represented by (\ref{eq:logistic_model}) implicitly assumes that the day of the week (DoW) and the week within the year are known exactly for each event/control location, $i$. Then, $DoW(i)$ and $w(i)$ are two known values and the corresponding fixed and random effects can be estimated. If the exact date of occurrence of event $i$ is unknown, $DoW(i)$ is unknown, while $w(i)$ is also unknown unless the \textit{from date} and the \textit{to date} belong to the same week of the calendar. In this scenario, one possibility is to discard all the events with an unknown date for the analysis. This avoids dealing with missing data and corresponds to a complete case analysis (considering only the data records with no missing values). Removing all these observations from the analysis leads to a reduction in both precision and power, which is undesired.

Hence, to include all these temporally-uncertain events within the modeling framework, one can treat each missing date as a random variable, as usually done in the context of Bayesian statistics to deal with missing data. Specifically, following the aoristic approach, a uniform prior is assigned to each date, considering the information provided by the \textit{from date} and \textit{to date} variables available, that is, $t_{i}^{event} \sim U(t_{i}^{from} - 0.5,t_{i}^{to} + 0.5)$. Then, in each iteration of the Monte Carlo Markov Chain (MCMC) process, a numeric date is sampled according to this distribution, which is rounded (to the nearest integer) to allow the computation of $DoW(i)$ and $w(i)$, and hence the consideration of all the data available in the estimation of the fixed and random effects of the model. For known dates, since $t_{to}=t_{from}$, all the sampled values for $t_{i}^{event}$ coincide with the exact (known) date. Moreover, note that subtracting and adding 0.5, respectively, to $t_{i}^{from}$ and $t_{i}^{to}$ is necessary to avoid reducing the weight of the extremes of the interval corresponding to the event in the prior distribution.

Thus, by treating $t_{i}^{event}$ as a random variable, no data is discarded for the analysis. In the remainder of the paper, this model is called the full model. In addition, we can also compute the posterior distribution $p(t_{i}^{event}=t|D)$, where $D$ from now on stands for the dataset used to fit the model, for $t\in\{t_{i}^{from},t_{i}^{from}+1,...,t_{i}^{to}-1,t_{i}^{to}\}$, to estimate the probability that a temporally-uncertain event has occurred in each of the dates within its associated interval of occurrence delimited by the Police.

The logistic regression models described in previous lines have been implemented in the NIMBLE system for Bayesian inference \citep{de2017programming}, based on MCMC procedures. 

\subsection{Model criticism}

In Bayesian analysis, model assessment and comparison is typically performed through some well-known goodness-of-fit measures such as the Deviance Information Criterion (DIC) introduced by \cite{spiegelhalter2002bayesian}, or the Watanabe–Akaike Information Criterion (WAIC) proposed by \cite{watanabe2010asymptotic}. However, these metrics are only useful to compare models with the same likelihood function, so the complete cases model and the full model described above cannot be compared in terms of these metrics (the likelihood functions differ since each model is fitted to a different dataset).

Therefore, a different strategy is needed. One possibility is to perform model criticism through the analysis of the distribution of the point estimates of the $\pi_i$'s, denoted by $\hat{\pi}_i$'s, each of which has been computed as the mean of the posterior distribution $p(\pi_i|D)$. Specifically, we first compare the distribution of the $\hat{\pi}_i$'s across models and location types (case vs. control). This enables us to appreciate if each model can discriminate between cases and controls and if we can find any remarkable difference. Second, considering the full model, it is examined if the distributions of the $\hat{\pi}_i$'s corresponding to certain and temporally-uncertain events differ.

In addition, it is also of interest to study the quality of the models as classification tools. For goodness-of-fit purposes, we can employ the in-sample predictive capability of a model. In this study, the F1 score and the Matthews correlation coefficient (MCC) \citep{matthews1975comparison} have been chosen for evaluating the in-sample predictive quality of the models. Other well-known metrics, such as accuracy, have been discarded for the analysis as they are unreliable for imbalanced datasets. Indeed, they tend to provide an overoptimistic estimation of the classifier ability on the majority class \citep{chicco2020advantages}. The F1 score is defined as follows:
$$\mathrm{F1}=\frac{2\mathrm{TP}}{2\mathrm{TP}+\mathrm{FP}+\mathrm{FN}}$$

Recent research has shown that the MCC performs better than other more typical metrics \citep{chicco2020advantages}. The MCC ranges from -1 (worst value) to 1 (best value) and is defined as follows:
$$\mathrm{MCC}=\frac{\mathrm{TP} \cdot \mathrm{TN} - \mathrm{FP} \cdot \mathrm{FN}}{\sqrt{(\mathrm{TP}+\mathrm{FP})\cdot(\mathrm{TP}+\mathrm{FN})\cdot(\mathrm{TN}+\mathrm{FP})\cdot(\mathrm{TN}+\mathrm{FN})}}\mathrm{,}$$
where TP is the number of true positives (positive prediction and actual event occurrence), FP the number of false positives (positive prediction but no event occurrence), FN the number of false negatives (negative prediction but actual event occurrence), and TN the number of true negatives (negative prediction and no event occurrence). In order to label an observation as a positive or a negative, a cutoff probability, $c\in]0,1[$, needs to be used as a threshold. Then, if $\hat{\pi}_i>c$ the observation is classified as a positive (burglary), whereas if $\hat{\pi}_i \leq c$, the observation is classified as a negative (no burglary). If we were working with a balanced dataset (with the same number of cases as controls), $c=0.5$ would be a natural choice. However, the datasets under analysis are imbalanced in favor of controls (by construction, they present a 5:1 ratio of controls to cases). Hence, lower values of $c$ are more suitable, otherwise, most of the observations will be classified as negatives and the classification would be far from optimal. For this reason, an analysis of the MCC as a function of $c$ is performed.

The formula of the MCC provides a point estimate of the predictive quality of the model. Following the approach proposed by \cite{gilardi2022multivariate} for the analysis of the balanced accuracy of a model, the distribution of the MCC has been estimated in this case. Specifically, the sampled values from the posterior distribution $p(\pi_i|D)$ can be used to simulate MCC values and therefore derive the distribution of the MCC, which allows a more complete comparison of the predictive ability of the two models fitted.


\subsection{Software}

The R programming language \citep{teamR} has been used for the analysis. In particular, the R packages \textsf{ggplot2} \citep{ggplot2}, \textsf{lubridate} \citep{grolemund2011dates}, \textsf{nimble} \citep{de2017programming}, \textsf{rgdal} \citep{rgdal}, \textsf{rgeos} \citep{rgeos}, \textsf{spatstat} \citep{baddeley2015spatial}, and \textsf{spdep} \citep{bivand2008applied} have been used.

\section{Results}
\label{Results}

\subsection{A simulation study}

Before proceeding to the analysis of the dataset of residential burglaries in Valencia, a simulation study is carried out to test the suitability of the proposed model under certain assumptions about the data. Specifically, we select the control observations that will also be used for the analysis of real data, which are extended with a sample of 3000 elements that will play the role of cases. This sample is obtained by resampling over the set of controls, so that the event of interest presents a constant intensity over the days of the week, and a constant spatio-temporal intensity (conditional on the population size by borough). Temporal uncertainty is introduced for a proportion of these cases following two different mechanisms that allow us to reflect that the presence of interval-censored observations may depend on temporal factors. Specifically, we simulate three datasets in which the probability of interval-censored observation depends on the day of the week (scenarios 1, 2 and 3), and three other datasets in which it depends on the week of the year (scenarios 4, 5 and 6). The objective is to verify whether the full model is able to deal with this uncertainty and recover the true baseline risk, which is constant in space and time as we have mentioned above. In order to mimic the basic characteristics of the real burglary dataset to be analyzed later, the magnitude of the temporal uncertainty (in days) for each of these events is obtained by simulating from an Exponential variable of rate $\lambda=0.2$ and by adding 1 to the simulated value (to ensure that the time uncertainty is at least of 1 day, otherwise there would be no uncertainty). Then, if $t_{i}$ is the actual temporal location of simulated event $i$, we compute $t_{i}^{from}=t_{i}-Round(u+1)$ and $t_{i}^{to}=t_{i}+Round(u+1)$, where $u\sim Exp(\lambda=0.2)$. The following paragraphs describe in detail how the interval-censored observations are chosen for each scenario.

Let $p_{DoW}$ be the probability that an event that actually occurs on day $DoW$ of the week presents some temporal uncertainty ($p_1$ corresponds to such probability for Mondays, $p_2$ for Tuesdays, and so on). We assume three scenarios for the $p_{DoW}$'s. In the first scenario, we assume equal probabilities within the week, that is, $p_{DoW}=0.4$, $DoW=1,...,7$. In the second and third scenarios, we assume that the probability that an event presents temporal uncertainty is greater if the event has actually occurred from Friday to Sunday. Particularly, we consider $\{p_1,p_2,p_3,p_4,p_5,p_6,p_7\}=\{0.3,0.3,0.3,0.3,0.5,0.5,0.5\}$ in scenario 2, and $\{p_1,p_2,p_3,p_4,p_5,p_6,p_7\}=\{0.2,0.2,0.2,0.2,0.6,0.6,0.6\}$ in scenario 3. Therefore, scenario 3 represents a more extreme scenario than scenario 2 in this regard.

Similarly, let $p_{w}$ be the probability that an event that actually occurs on week $w$ of the period 2016-2017 ($w=1,...,104$) presents temporal uncertainty. We also assume three scenarios for the $p_{w}$'s. In scenario 4, we assume $p_{w}=0.4$, $w=1,...,104$. In the scenarios 5 and 6, we assume that the probability that an observation is interval-censored is greater in the last quarter of the year, which corresponds to weeks 40 to 52, approximately. We denote $W_{Q4}=\{40,...,52,92,...,104\}$ the set of weeks in the last quarter of 2016 and 2017. Hence, we choose $p_{w}=0.5$ for $w\in W_{Q4}$ and $p_{w}=0.3$ for $w\notin W_{Q4}$ in scenario 5, and $p_{w}=0.6$ for $w\in W_{Q4}$ and $p_{w}=0.2$ for $w\notin W_{Q4}$ in scenario 6.

\begin{figure}[htbp]
 \centering
 \subfloat[]{\includegraphics[width=5cm,angle=-90]{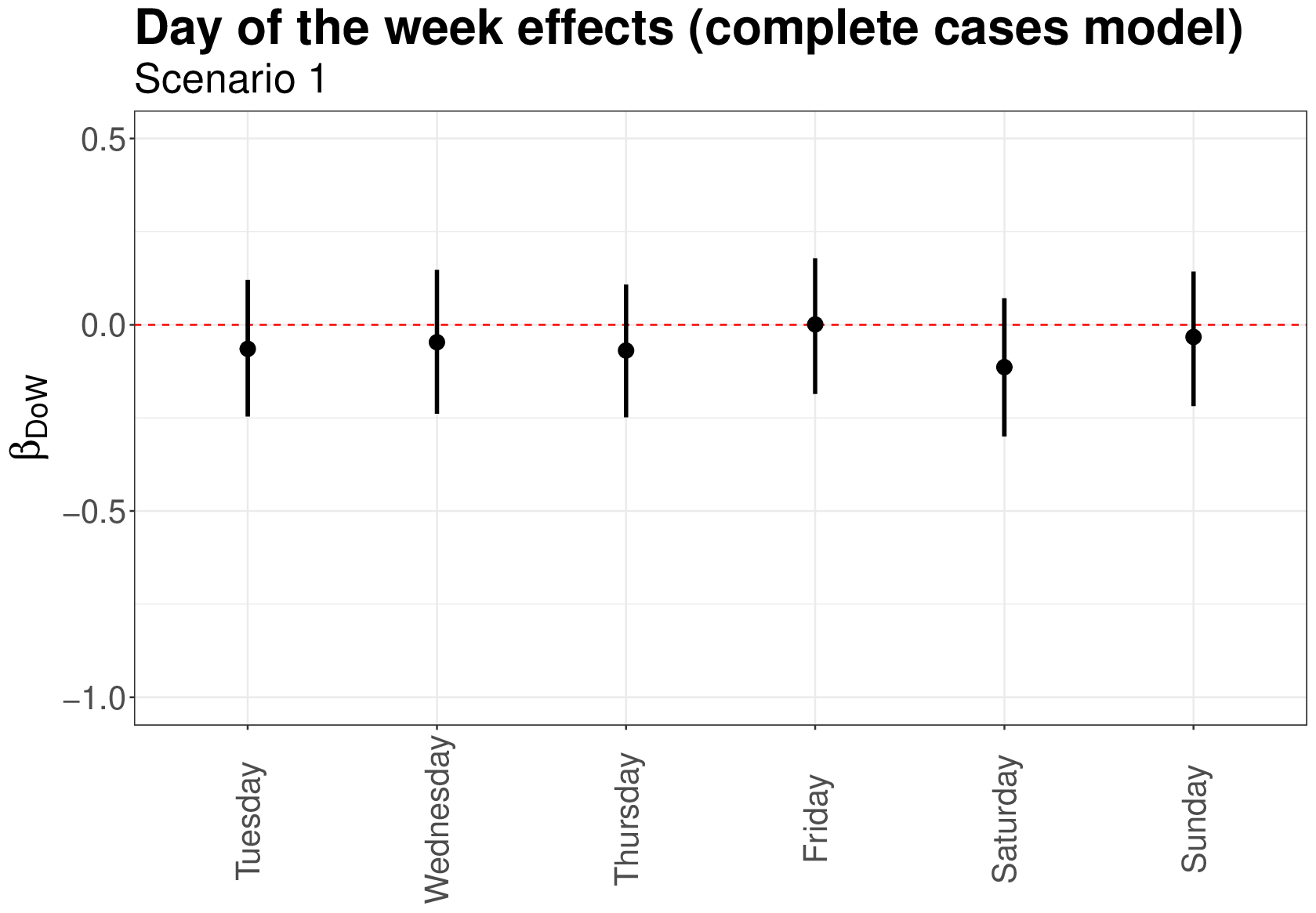}\label{fig:dow_effects_sim_a}}
 \subfloat[]{\includegraphics[width=5cm,angle=-90]{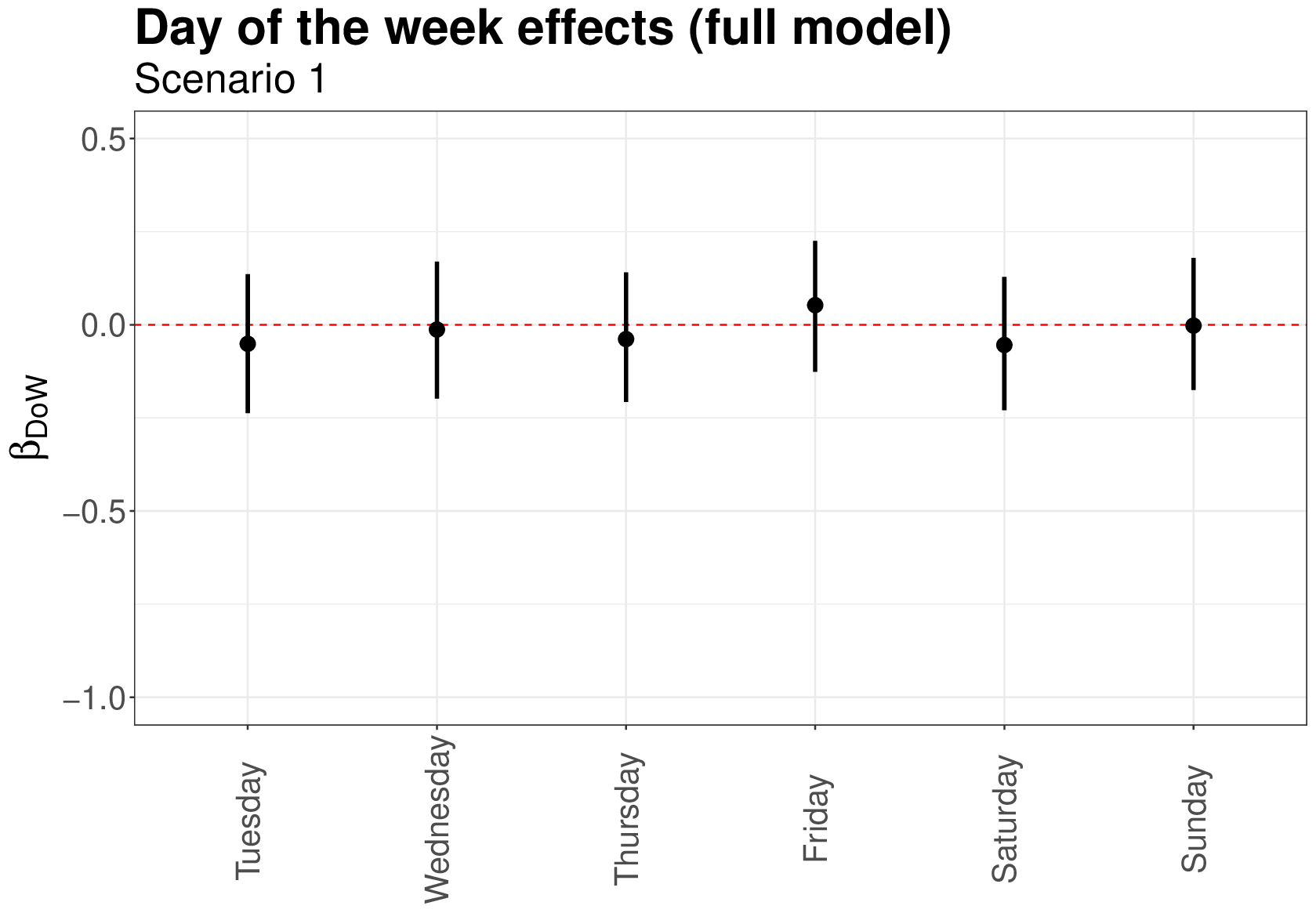}\label{fig:dow_effects_sim_b}}\\
 \subfloat[]{\includegraphics[width=5cm,angle=-90]{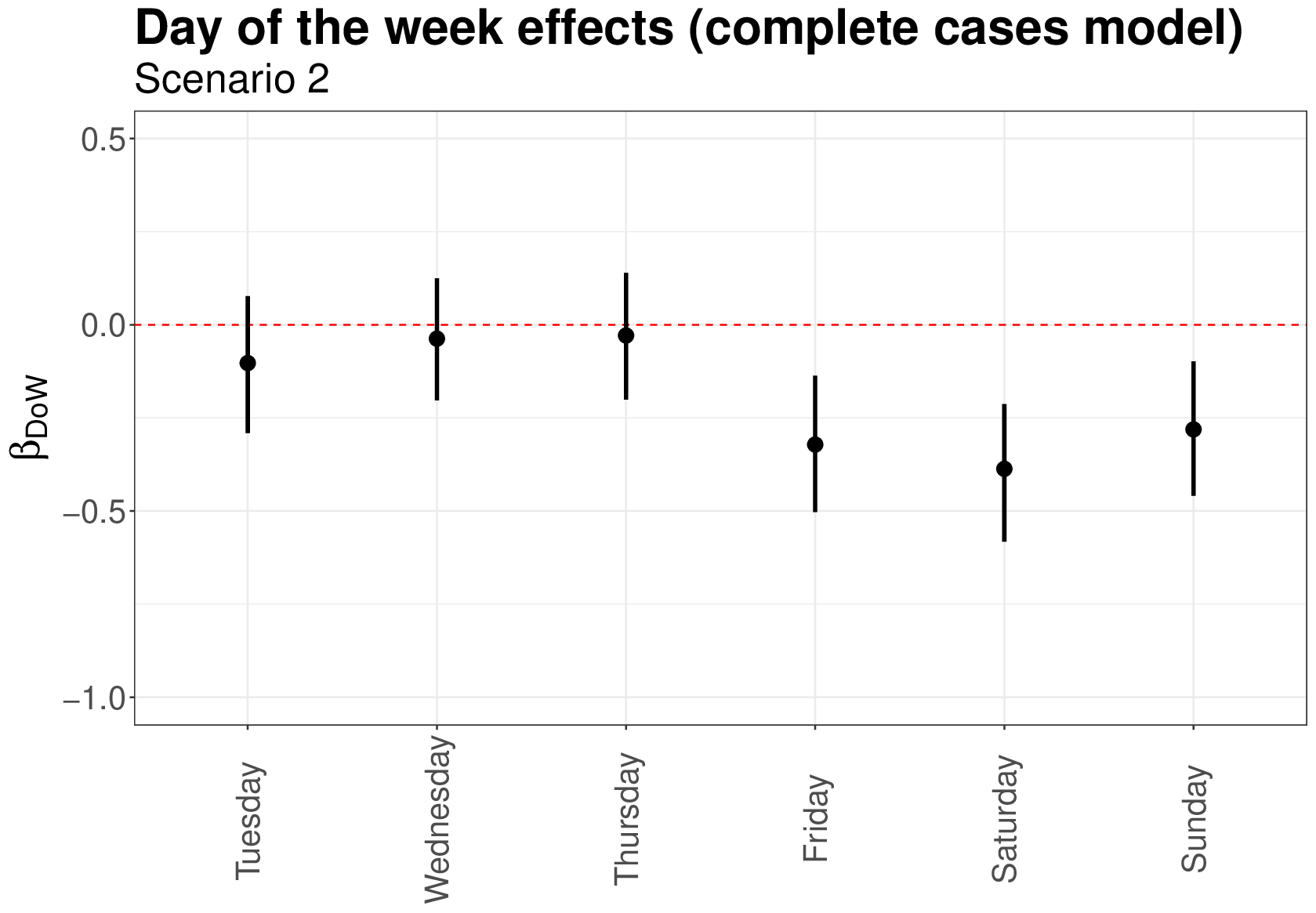}\label{fig:dow_effects_sim_c}}
 \subfloat[]{\includegraphics[width=5cm,angle=-90]{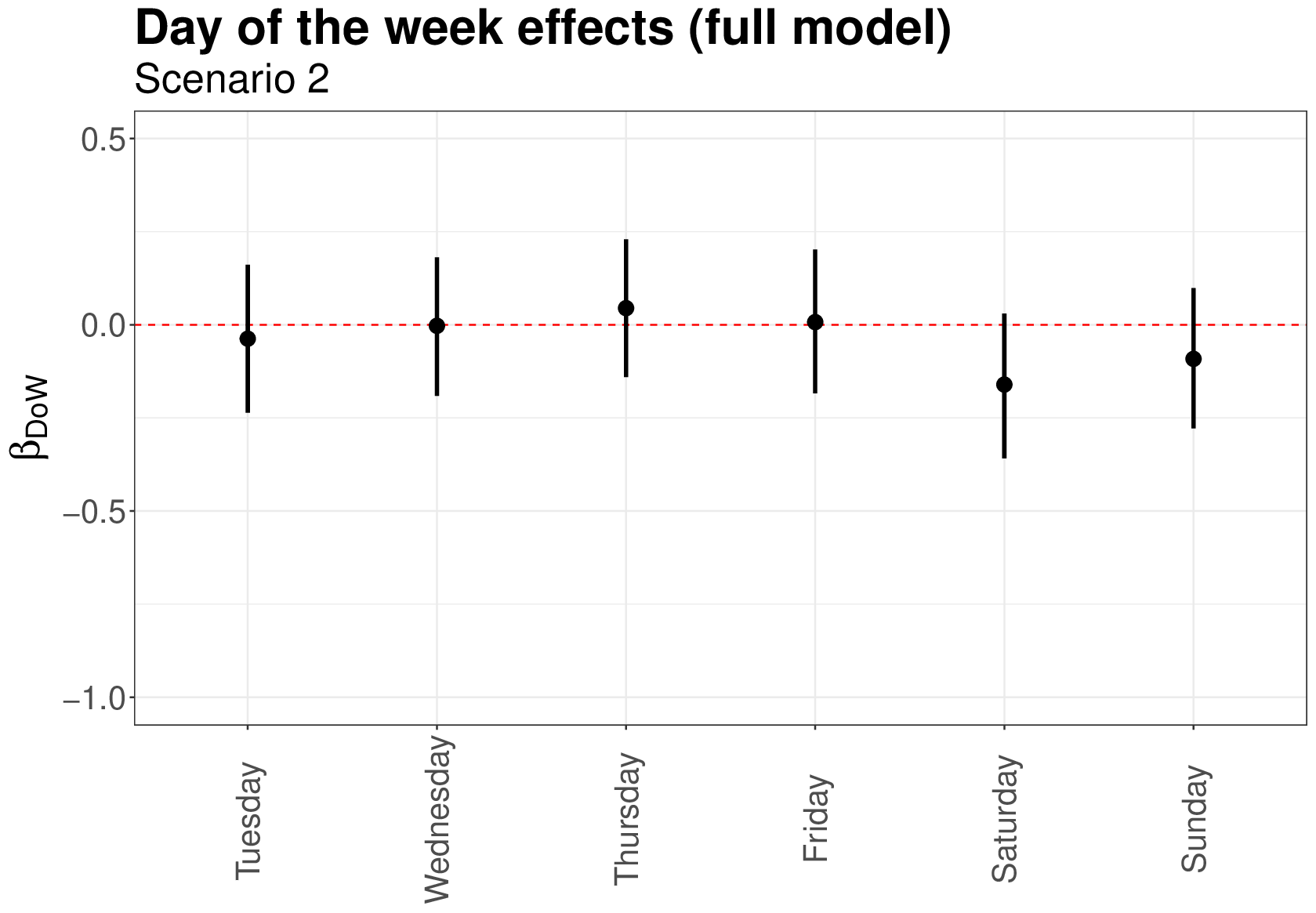}\label{fig:dow_effects_sim_d}}\\
 \subfloat[]{\includegraphics[width=5cm,angle=-90]{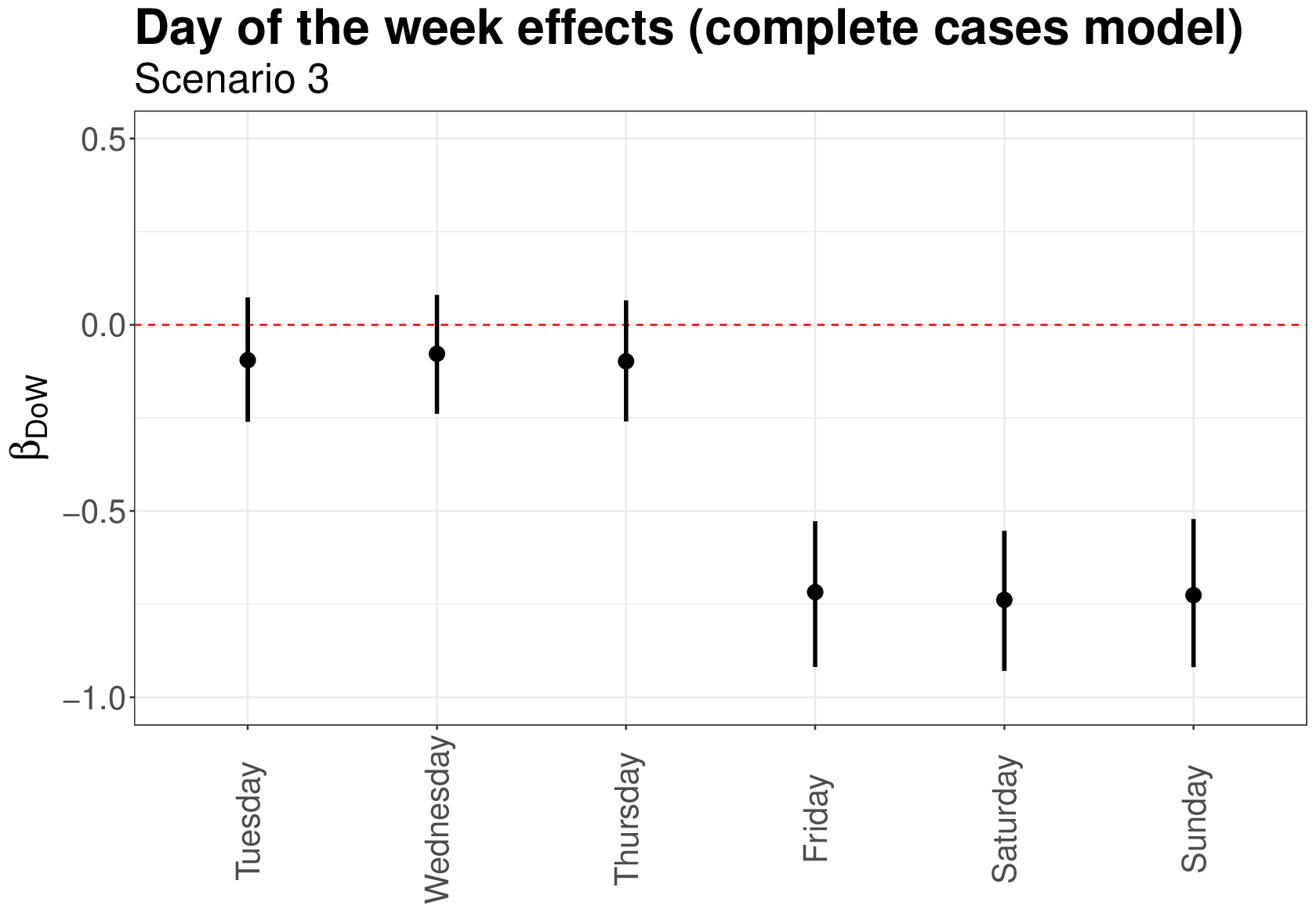}\label{fig:dow_effects_sim_e}}
 \subfloat[]{\includegraphics[width=5cm,angle=-90]{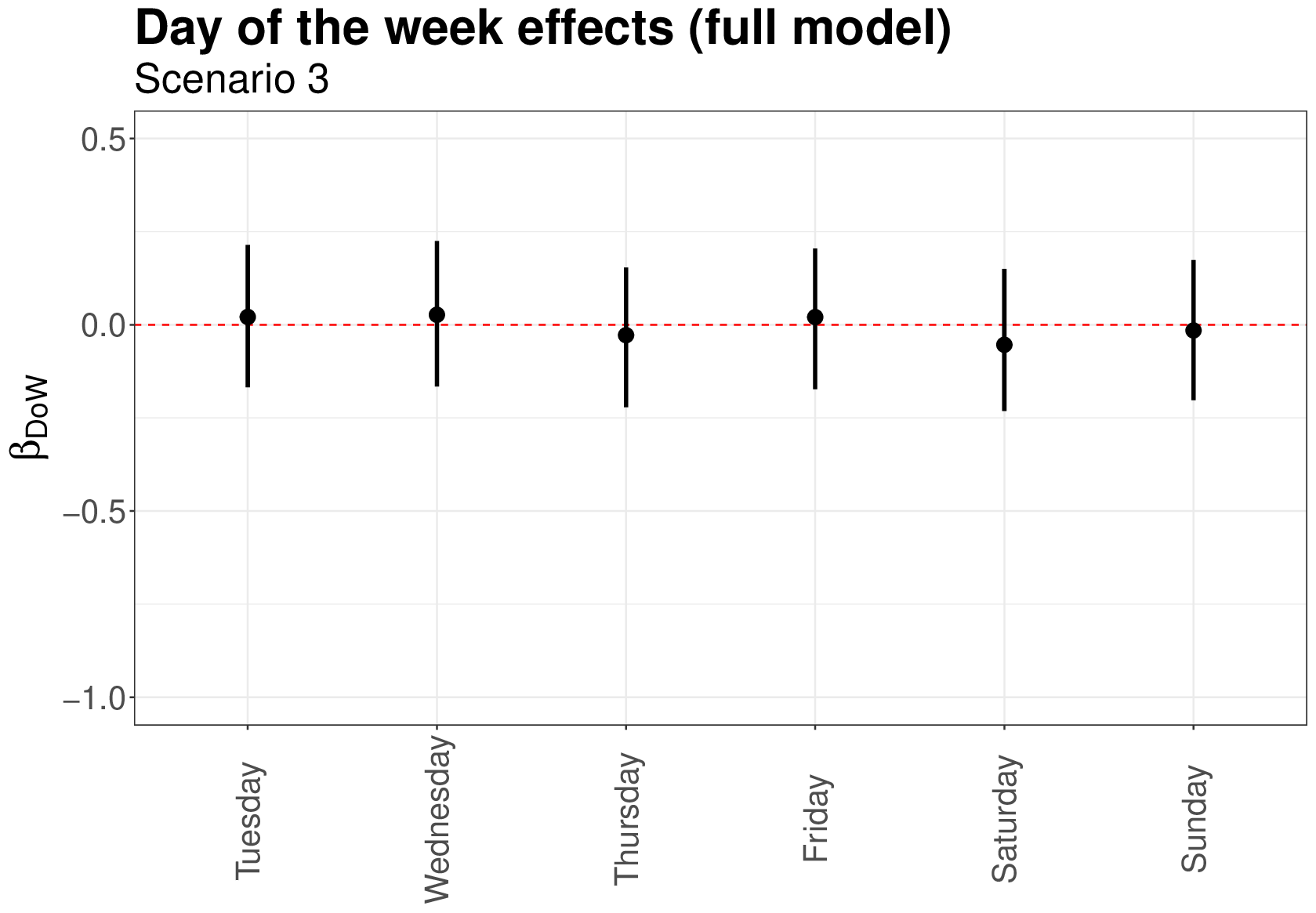}\label{fig:dow_effects_sim_f}}
\caption{Day of the week effect estimates yielded by the complete cases model and the full model under the simulated scenarios 1, 2, and 3}
\label{fig:dow_effects_sim}
\end{figure}

Figure \ref{fig:dow_effects_sim} shows the day of the week effects estimated under simulated scenarios 1, 2, and 3, considering the complete cases and the full model. The greater presence of temporally-uncertain observations from Friday to Sunday in scenarios 2 and 3 causes these days to be underrepresented in the dataset used for fitting the full model, as these interval-censored events are removed from the analysis. This results in lower estimates for Friday, Saturday and Sunday, which would lead to the erroneous conclusion that fewer events occur on these days. This does not occur in scenario 1, since in this case an equal proportion of events is removed for all days of the week, which makes the estimates of the effects similar. On the other hand, it can be seen that the full model, which deals with the temporal uncertainty existing in some observations, is able to correctly estimate that all days of the week roughly present the same effect.

\begin{figure}[htbp]
 \centering
 \subfloat[]{\includegraphics[width=5cm,angle=-90]{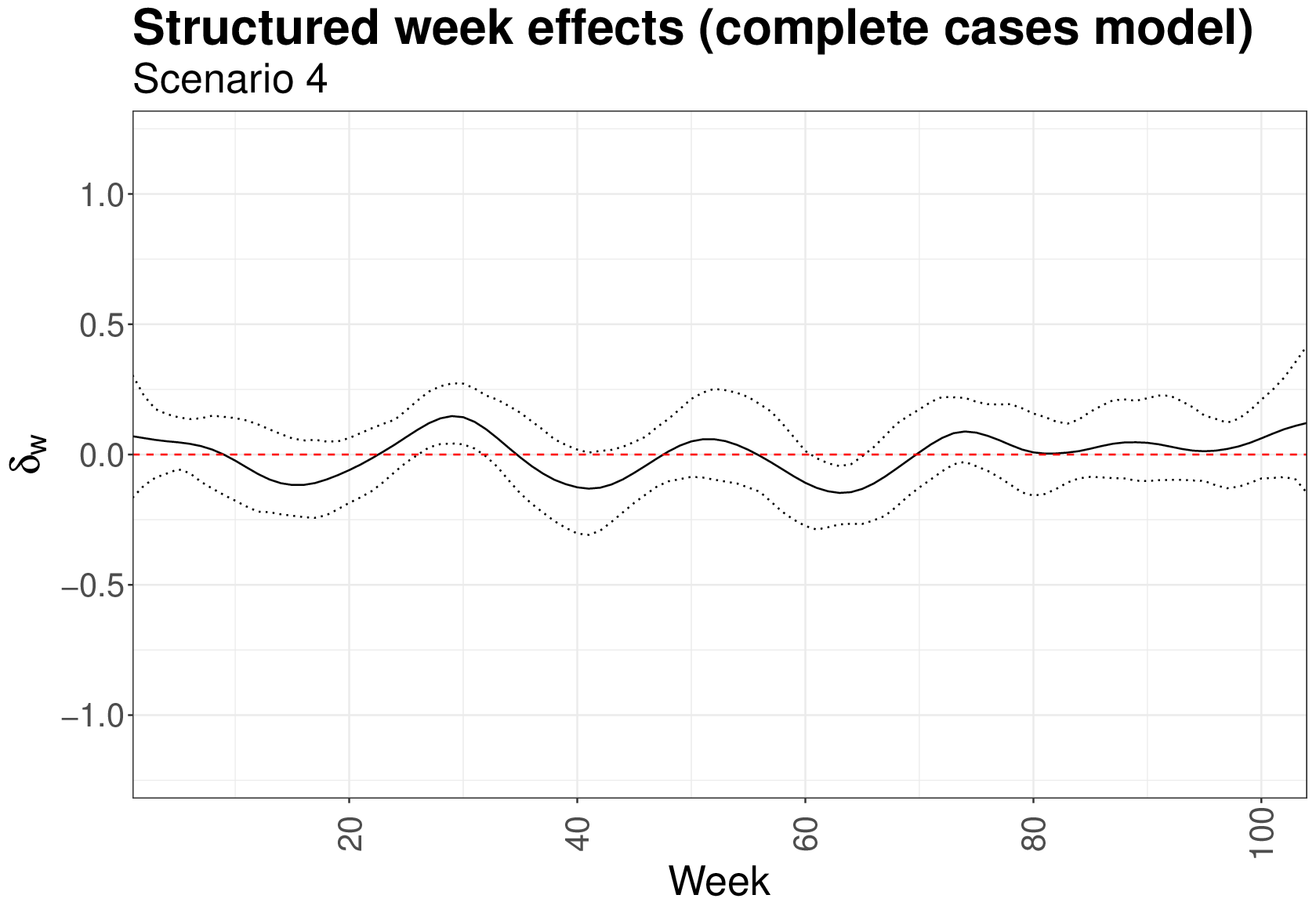}\label{fig:week_effects_sim_a}}
 \subfloat[]{\includegraphics[width=5cm,angle=-90]{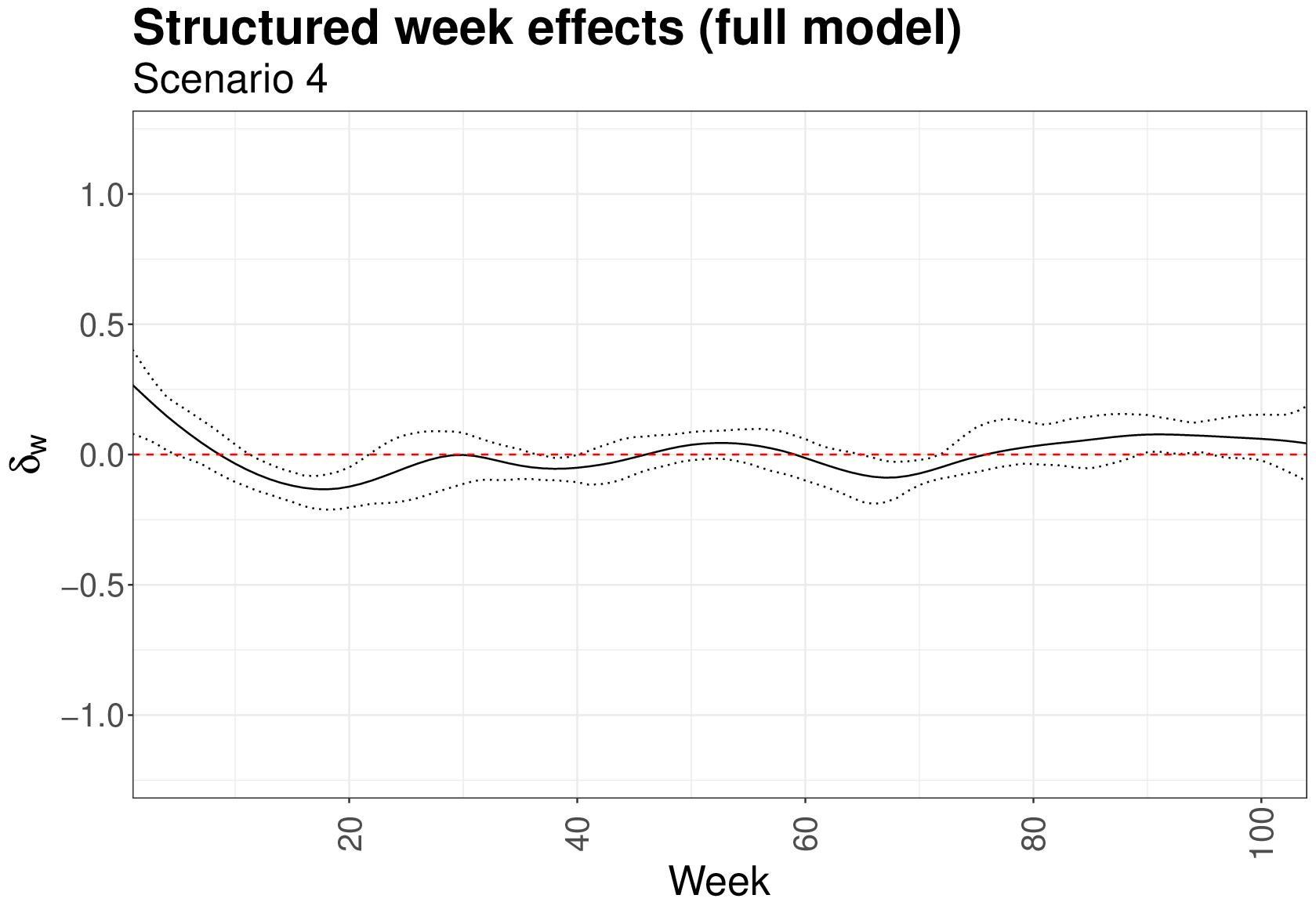}\label{fig:week_effects_sim_b}}\\
 \subfloat[]{\includegraphics[width=5cm,angle=-90]{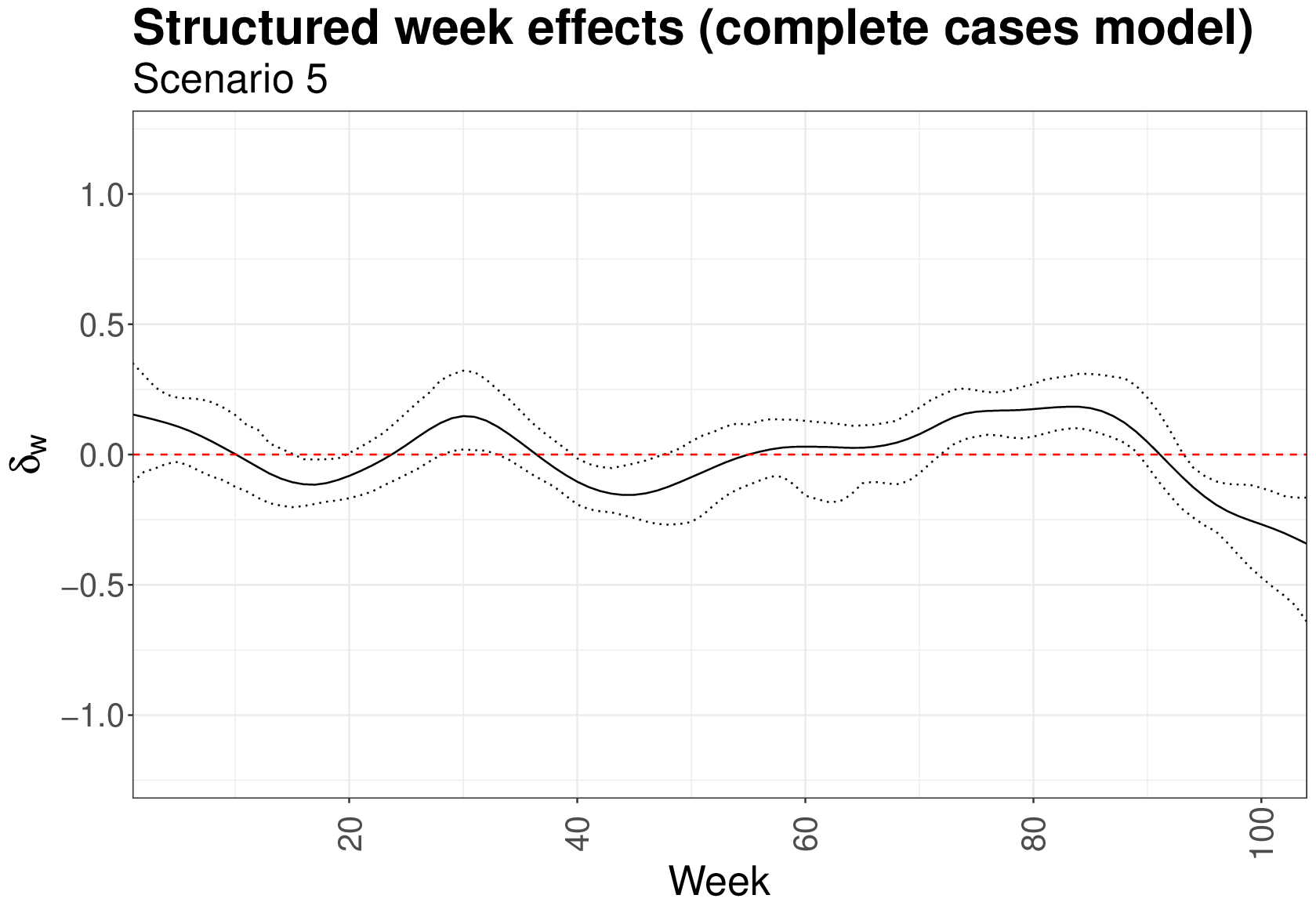}\label{fig:week_effects_sim_c}}
 \subfloat[]{\includegraphics[width=5cm,angle=-90]{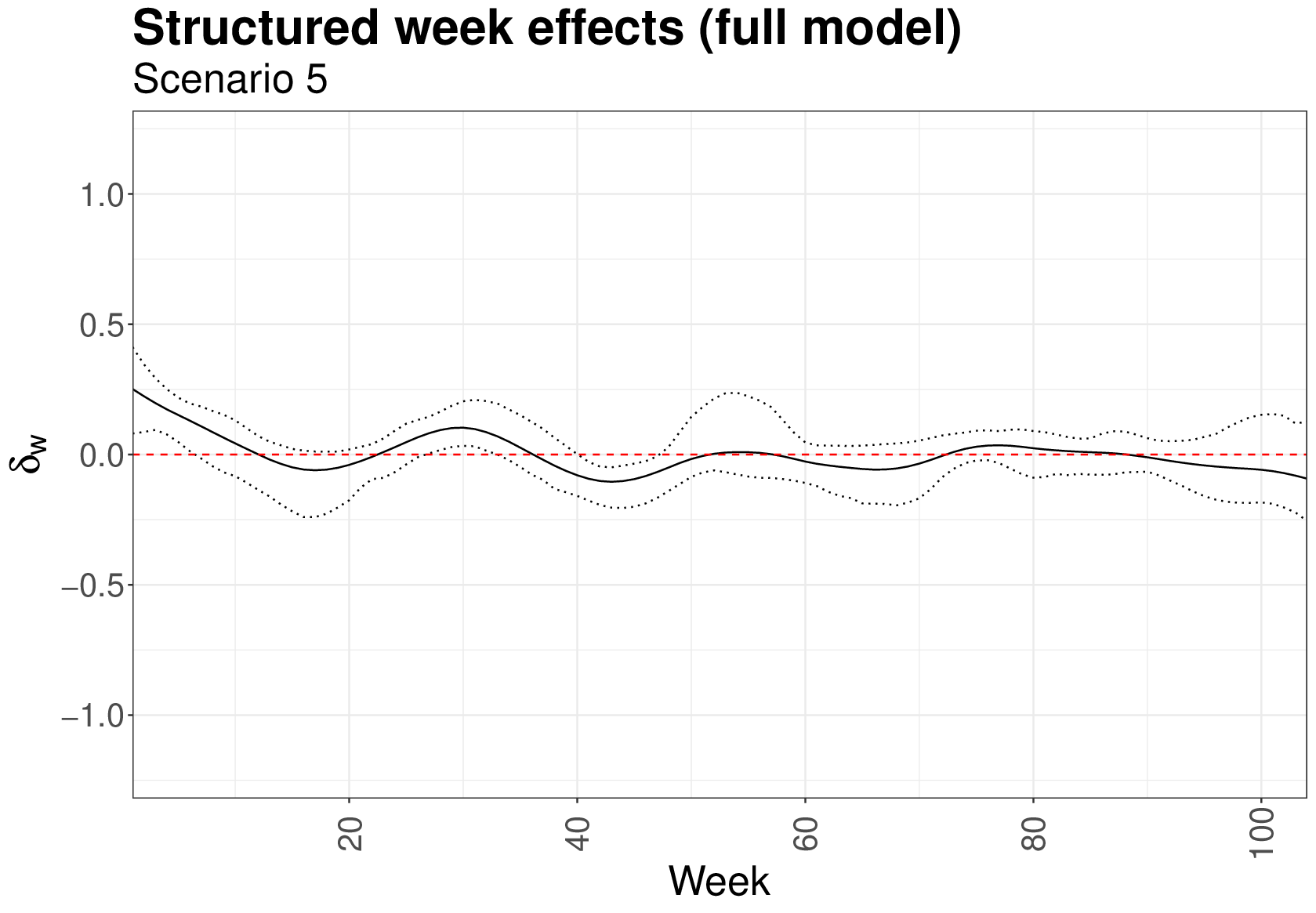}\label{fig:week_effects_sim_d}}\\
 \subfloat[]{\includegraphics[width=5cm,angle=-90]{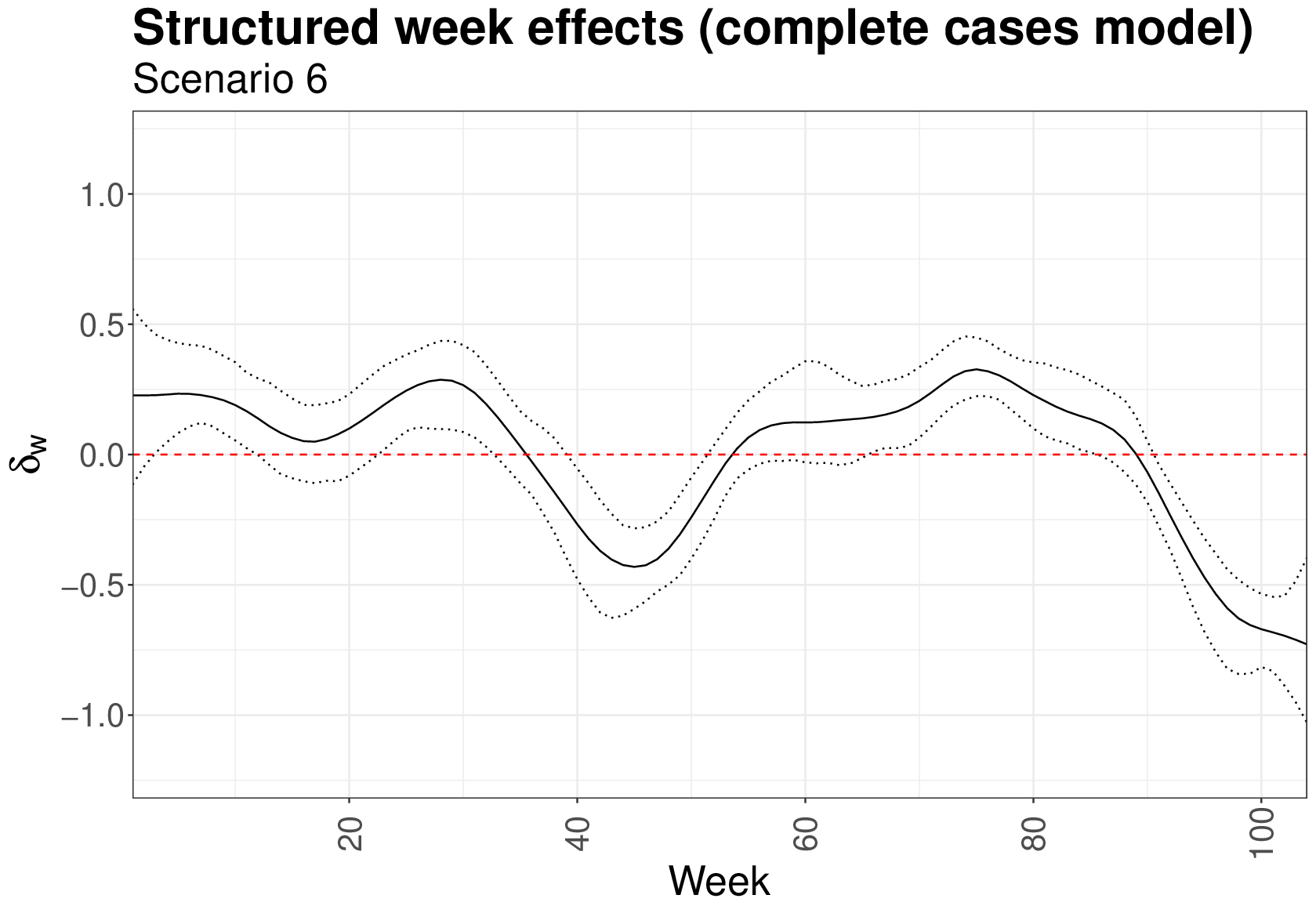}\label{fig:week_effects_sim_e}}
 \subfloat[]{\includegraphics[width=5cm,angle=-90]{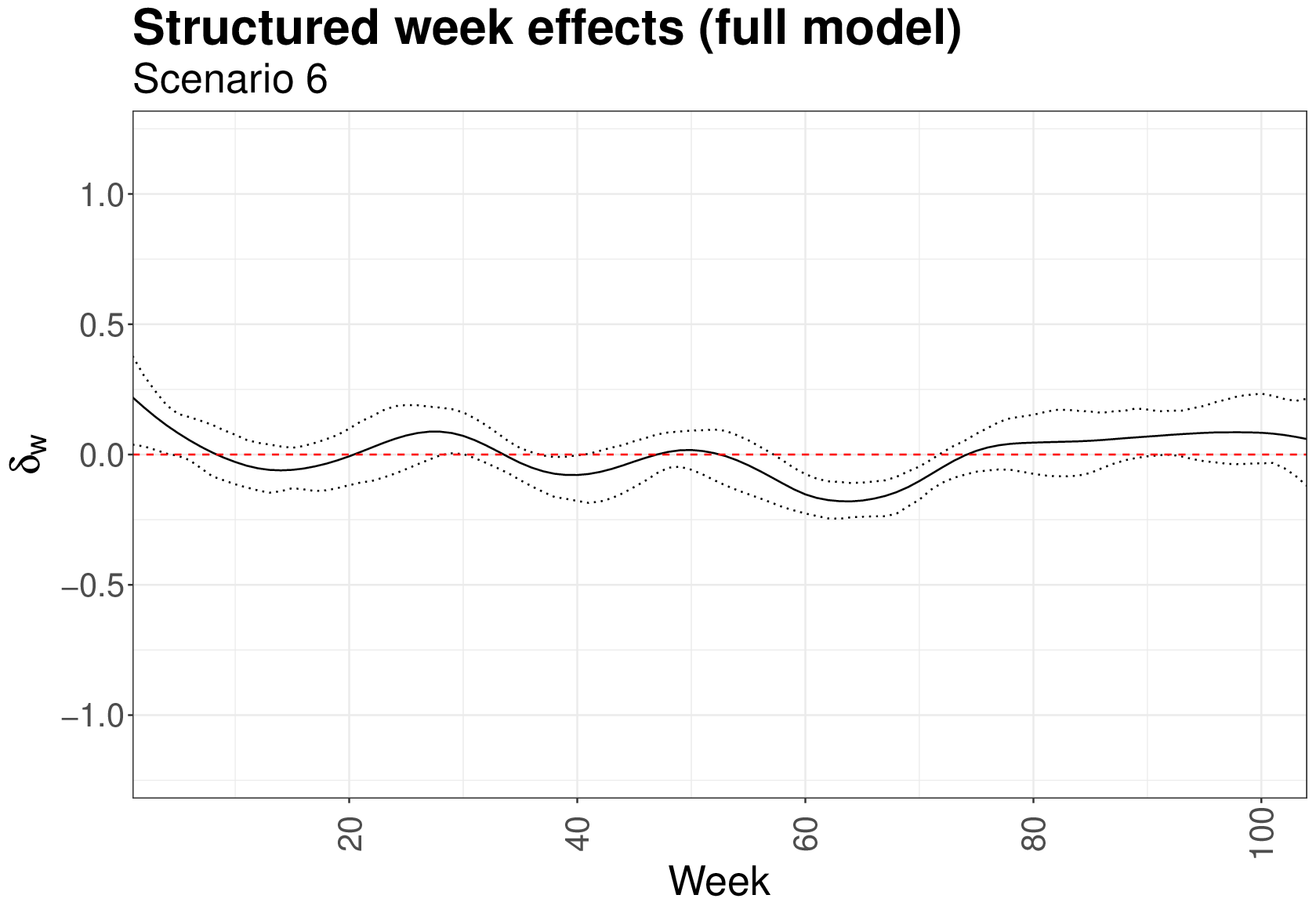}\label{fig:week_effects_sim_f}}
\caption{Day of the week effect estimates yielded by the complete cases model and the full model under the simulated scenarios 4, 5, and 6}
\label{fig:week_effects_sim}
\end{figure}

Figure \ref{fig:week_effects_sim} shows the structured week effects estimated under simulated scenarios 4, 5, and 6, through both the complete cases and the full model. As expected, the complete cases model detects some weeks in the last quarter of the year as being of lower risk, due to the higher proportion of temporally-uncertain events simulated for this part of the calendar. In contrast, the full model returns an estimate of the structured temporal random effect that oscillates around 0, consistent with the absence of a week effect. Therefore, the full model allows capturing the underlying level of risk despite the presence of temporally-uncertain observations.

\subsection{A real data analysis}

\subsubsection{Complete cases analysis vs. full analysis}

A major objective of the article is to compare the results derived from using the complete cases model (discarding temporally-uncertain events) and the full model, considering the burglary dataset recorded in Valencia during the years 2016 and 2017. Specifically, the goal is to check whether the complete cases model gives rise to biased estimates of the parameters and whether the full model can capture the real temporal distribution of the events. Thus, Table \ref{table_models} summarizes the results in terms of the point and interval estimates of the parameters involved in the two models, allowing direct comparison. First, we note that the estimate of $\alpha$ is smaller for the complete cases model. This is a consequence of the fact that the dataset considered for fitting this model has a smaller proportion of cases than the one used for the full model, which makes the estimate of the baseline probability of event occurrence also lower. In any case, the $\alpha$ parameter has no major relevance in terms of interpretation.

\begin{table}[ht]
\centering
\begin{tabular}{lrrr|lrrr}
\hline
\multicolumn{4}{c|}{\textbf{Complete cases model}} & \multicolumn{4}{c}{\textbf{Full model}} \\ 
  \hline
\textbf{Parameter} & \textbf{Est}. & \textbf{Lo.} & \textbf{Up.} & \textbf{Parameter} & \textbf{Est.} & \textbf{Lo.} & \textbf{Up.} \\ 
  \hline
$\alpha$ & -2.17 & -2.34 & -2.01 & $\alpha$ & -1.74 & -1.89 & -1.60 \\ 
$\beta_{Tuesday}$ & -0.04 & -0.25 & 0.17 & $\beta_{Tuesday}$ & -0.15 & -0.34 & 0.04 \\ 
$\beta_{Wednesday}$ & 0.13 & -0.06 & 0.33 & $\beta_{Wednesday}$ & 0.00 & -0.17 & 0.17 \\ 
$\beta_{Thursday}$ & -0.04 & -0.24 & 0.16 & $\beta_{Thursday}$ & -0.10 & -0.28 & 0.09 \\ 
$\beta_{Friday}$ & 0.01 & -0.18 & 0.22 & $\beta_{Friday}$ & 0.16 & -0.02 & 0.34 \\ 
$\beta_{Saturday}$ & -0.07 & -0.27 & 0.12 & $\beta_{Saturday}$ & 0.18 & 0.00 & 0.36 \\ 
$\beta_{Sunday}$ & -0.30 & -0.50 & -0.09 & $\beta_{Sunday}$ & -0.04 & -0.23 & 0.16 \\ 
$\tau_{\delta}$ & 1748.60 & 543.62 & 3924.02 & $\tau_{\delta}$ & 880.02 & 187.52 & 2172.18 \\ 
$\tau_{\varepsilon}$ & 35.99 & 13.94 & 91.83 & $\tau_{\varepsilon}$ & 31.35 & 12.72 & 75.96 \\ 
$\tau_{u}$ & 2.33 & 1.21 & 4.33 & $\tau_{u}$ & 2.11 & 1.18 & 3.60 \\ 
$\tau_{v}$ & 111.07 & 12.66 & 529.44 & $\tau_{v}$ & 181.80 & 13.90 & 1036.27 \\ 
   \hline
\end{tabular}
\caption{Parameter estimates (Est.) yielded by the two models considered, along with the upper (Up.) and lower (Lo.) bound of the 95\% credibility intervals associated with each parameter. Each $\tau$ parameter corresponds to the precision of the corresponding random effect, being defined as the inverse of its variance. For instance, $\tau_{\delta}=1/\sigma^{2}_{\delta}$ is the precision of the structured temporal random effect}
\label{table_models}
\end{table}

\begin{figure}[htbp]
 \centering
 \subfloat[]{\includegraphics[width=5cm,angle=-90]{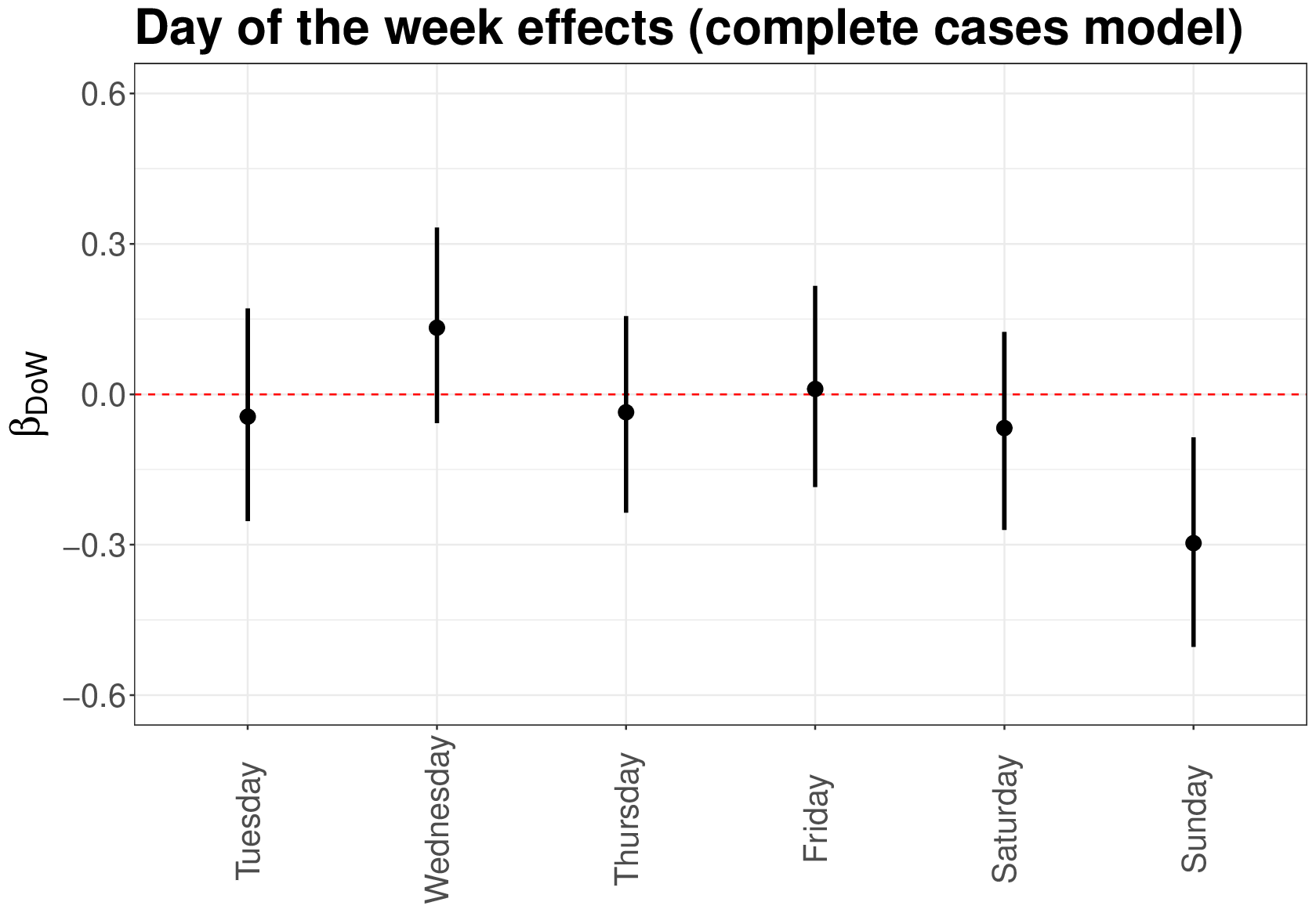}\label{fig:dow_effects_a}}
 \subfloat[]{\includegraphics[width=5cm,angle=-90]{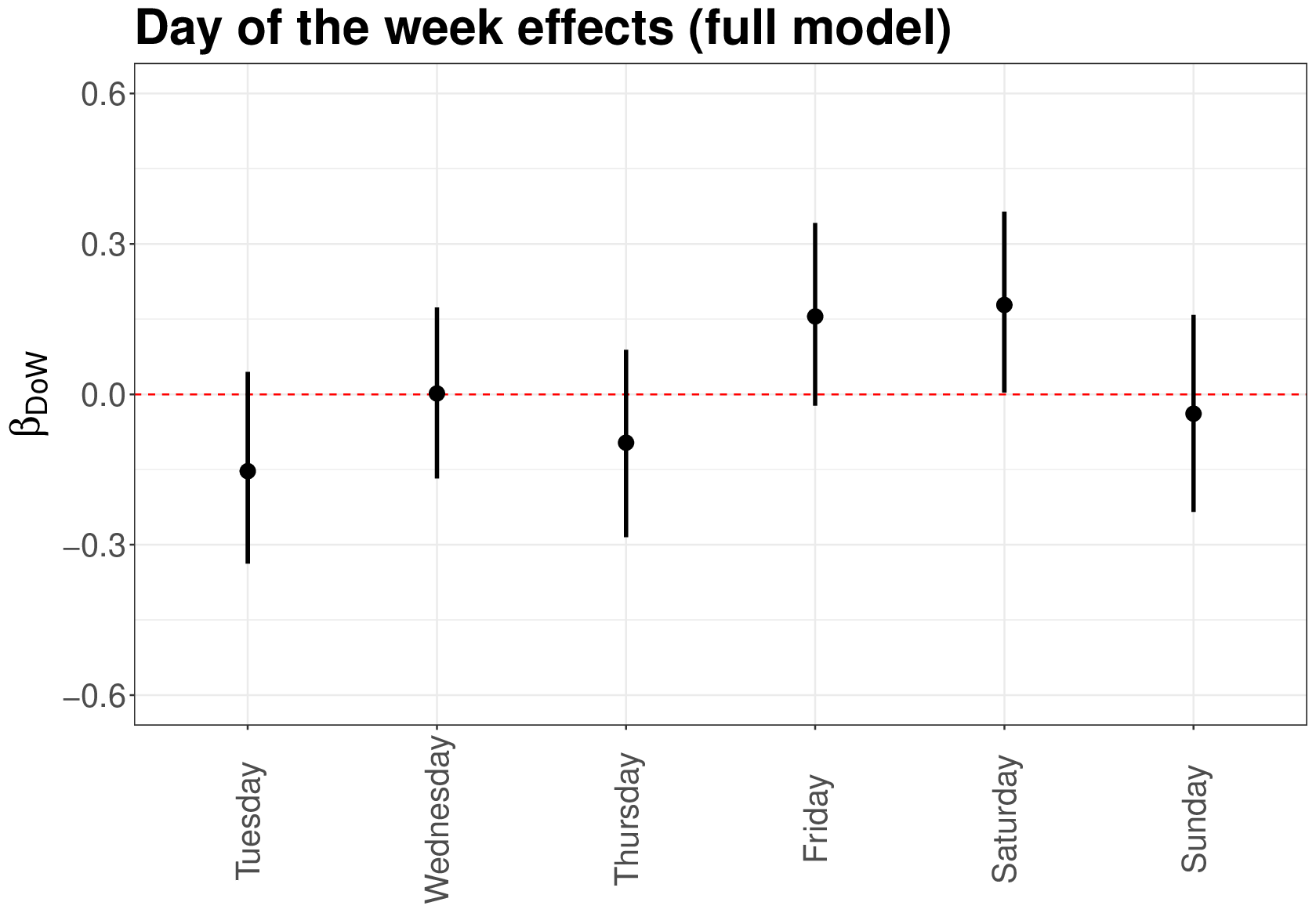}\label{fig:dow_effects_b}}\\
\caption{Day of the week effect estimates yielded by the complete cases model (a) and the full model (b)}
\label{fig:dow_effects}
\end{figure}

\begin{figure}[htbp]
 \centering
 \includegraphics[width=8cm,angle=-90]{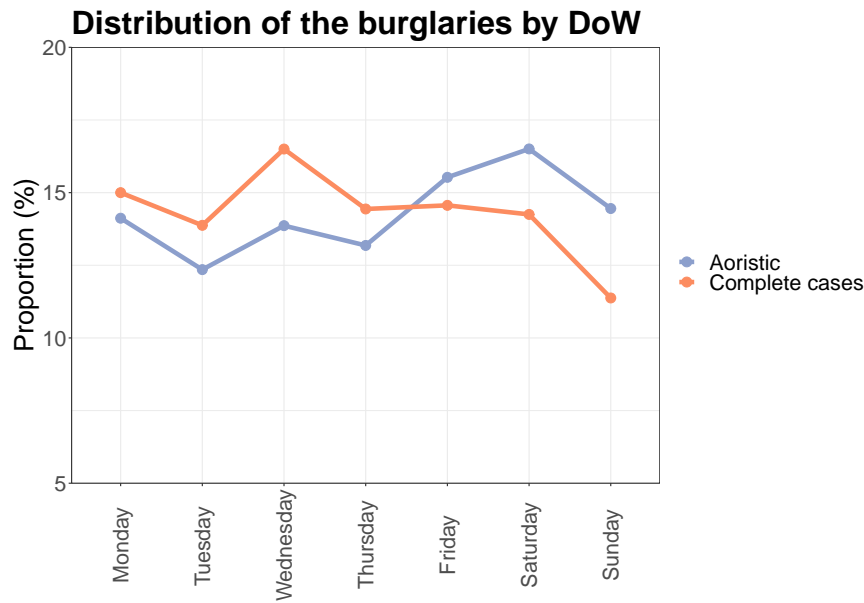}
\caption{Distribution of the burglaries by day of the week (DoW), considering only the events for which the exact date is known (complete cases analysis), and the whole available dataset by performing an aoristic analysis of the temporally-uncertain events}
\label{fig:aoristic_vs_exact}
\end{figure}

It is of greater interest to analyze the effects of the days of the week, as represented by the $\beta_{DoW}$ parameters (which represent the variation in risk in comparison to Mondays, the reference level). In this case, the differences are notable and of great relevance from a practical point of view. As shown in Figure \ref{fig:dow_effects}, according to the complete cases model, crime risk is notably lower on Sundays, whereas the highest estimate of crime risk corresponds to Wednesdays. In contrast, the full model yields Fridays and Saturdays as the two days with high burglary risk (even though the lower bound of the 95\% credibility interval associated with $\beta_{Friday}$ is slightly below 0), while Sundays do not display low burglary risk. Therefore, the results differ markedly depending on the model considered. In fact, the differences between models can be understood if an aoristic analysis of the distribution of residential burglaries by day of the week is performed, as shown in Figure \ref{fig:aoristic_vs_exact}. It can be observed how the fact of considering temporally-uncertain events in the analysis causes the proportion of crime to reach the highest values on the weekend, especially on Saturdays. In other words, this suggests that there is a higher proportion of temporally-uncertain events that cover (partially or totally) the weekend, possibly because on these days of the week part of the population stays in a second residence, or simply because of the changes in daily routines during the weekend, which could facilitate the action of burglars in certain time slots. This type of plausible assumption could be studied by considering an hour-level analysis, although it could also complicate the model estimation. In any case, the complete cases model entirely misses this type of information, which is recovered by the full model, despite the presence of interval-censored data.

\begin{figure}[htbp]
 \centering
  \subfloat[]{\includegraphics[width=5cm,angle=-90]{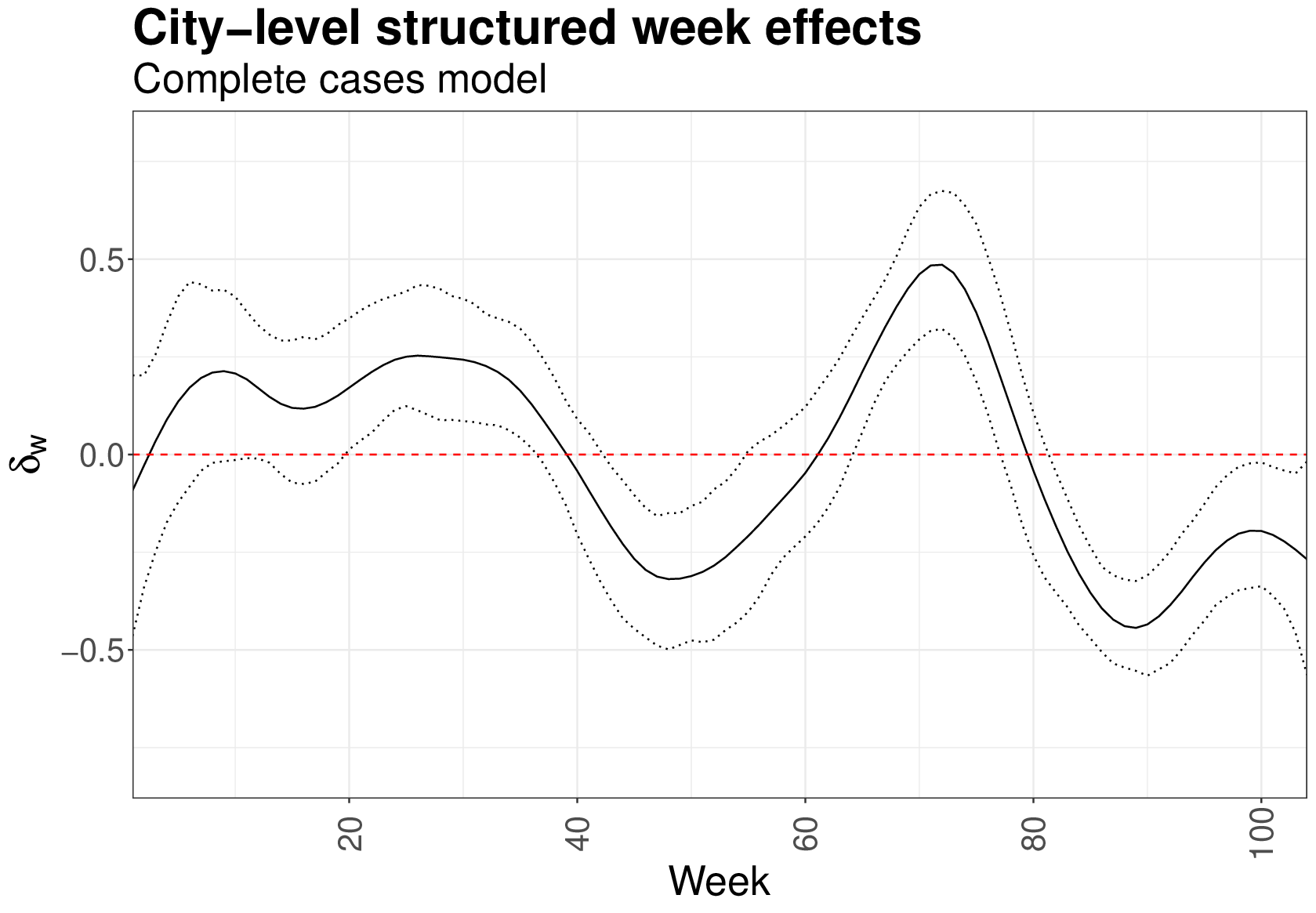}\label{fig:week_effects_str_impute_a}}
  \subfloat[]{\includegraphics[width=5cm,angle=-90]{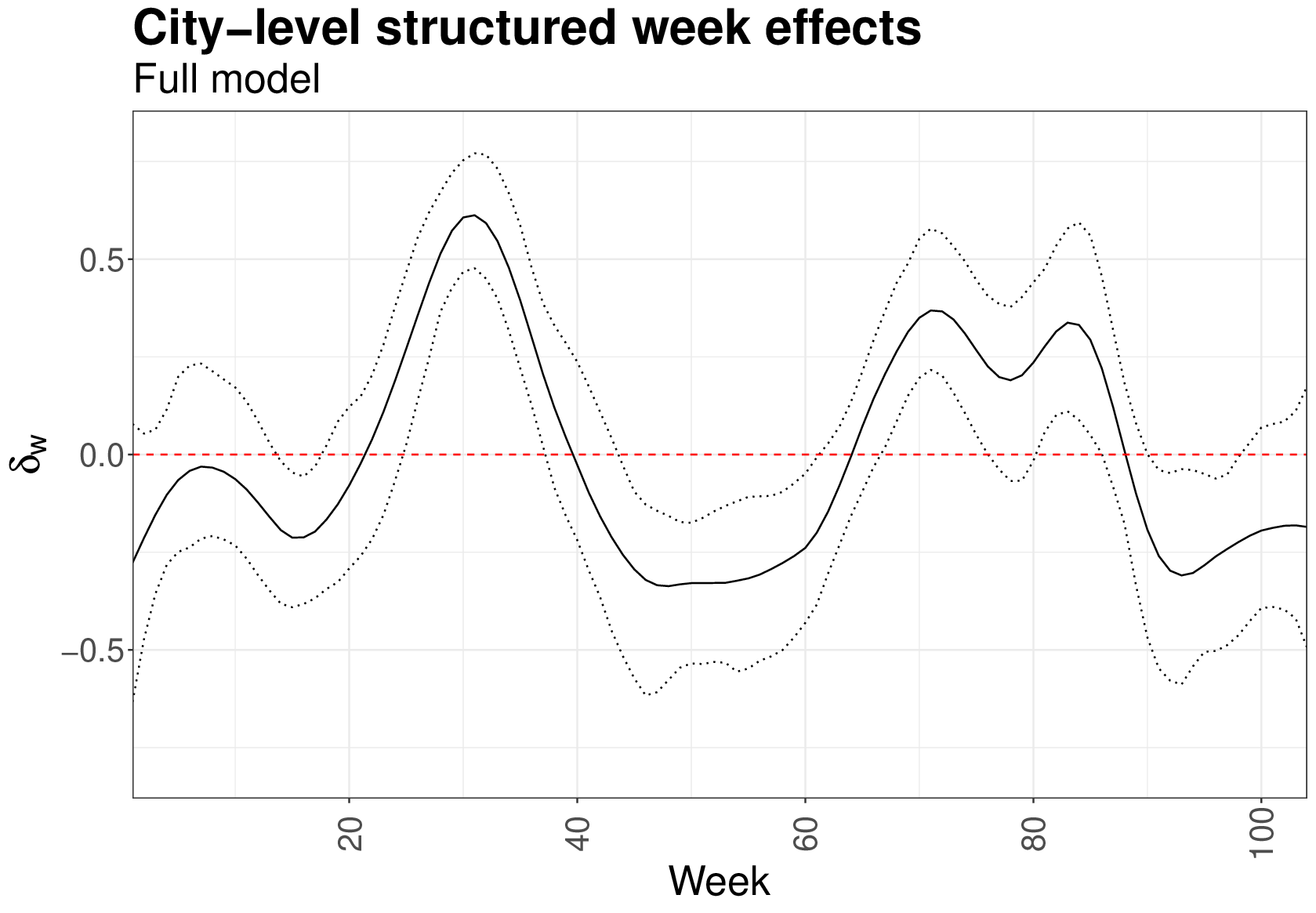}\label{fig:week_effects_str_impute_b}}
\caption{Week effect estimates (structured component) yielded by the complete cases model (a) and the full model (b). The dotted lines delimit the 95\% credibility interval associated with each point estimate}
\label{fig:week_effects_str_impute}
\end{figure}

Once the only fixed effect of the model (day of the week) has been analyzed, the estimates of the temporal and spatial random effects given by both models are compared. First, Figure \ref{fig:week_effects_str_impute} shows the estimates of the structured temporal random effect, $\delta_w$. Although the overall behavior of the temporal trend captured by this effect is similar for both models, certain differences arise. For instance, the complete cases model determines a peak in burglary risk at the beginning of the study period, which is not determined by the full model. Besides, the full model detects a double peak in crime risk around weeks 70 to 90, whereas the complete cases model locates a single peak within that period. Finally, the peak detected around week 30 by both models is notably higher in the estimation provided by the full model. Indeed, the full model can detect more variability at the week level than the complete cases model. This can be guessed from Figure \ref{fig:week_effects_str_impute}, but also by comparing the estimates of the precision of the random effect $\delta_w$. As shown in Table \ref{table_models}, the estimate of $\tau_{\delta}$ is notably smaller in the case of the full model, which confirms that the random effect $\delta_w$ captures more variability in the latter model, since $\tau_{\delta}=1/\sigma^2_{\delta}$.

\begin{figure}[htbp]
 \centering
 \includegraphics[width=8cm,angle=-90]{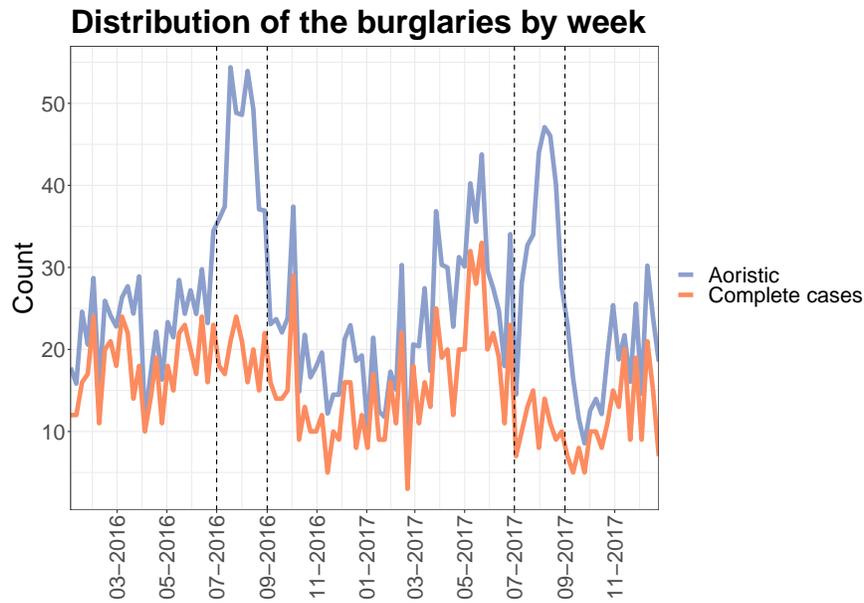}
\caption{Distribution of the burglaries by week, considering only the events for which the exact date is known (complete cases analysis), and the whole available dataset by performing an aoristic analysis of the temporally-uncertain events. The dashed lines delimit the summer months (July and August) of both years}
\label{fig:aoristic_vs_exact_week}
\end{figure}

The aoristic analysis of the distribution of residential burglaries by week allows us to verify, once again, that the full model adequately captures the temporal distribution of burglaries. Thus, as shown in Figure \ref{fig:aoristic_vs_exact_week}, the presence of temporally-uncertain events is notably higher in the summer months (July and August), when most residents enjoy holiday periods, increasing the likelihood that homes will be empty for days or even weeks. Besides, the aoristic analysis also reveals a peak in burglary counts during May 2017, which can be assumed to be the consequence of a separate process from the one corresponding to the summer peak. This peak in May 2017 is the only one detected by the complete cases model, since the proportion of temporally-uncertain events is quite low during this period, as shown in Figure \ref{fig:aoristic_vs_exact_week}. Paradoxically, the complete cases model indicates that the summer period of 2017 is a low-risk period, but this is only the consequence of data underrepresentation because of the presence of many temporally-uncertain events within these summer months.

\begin{figure}[htbp]
 \centering
 \subfloat[]{\includegraphics[width=5cm,angle=-90]{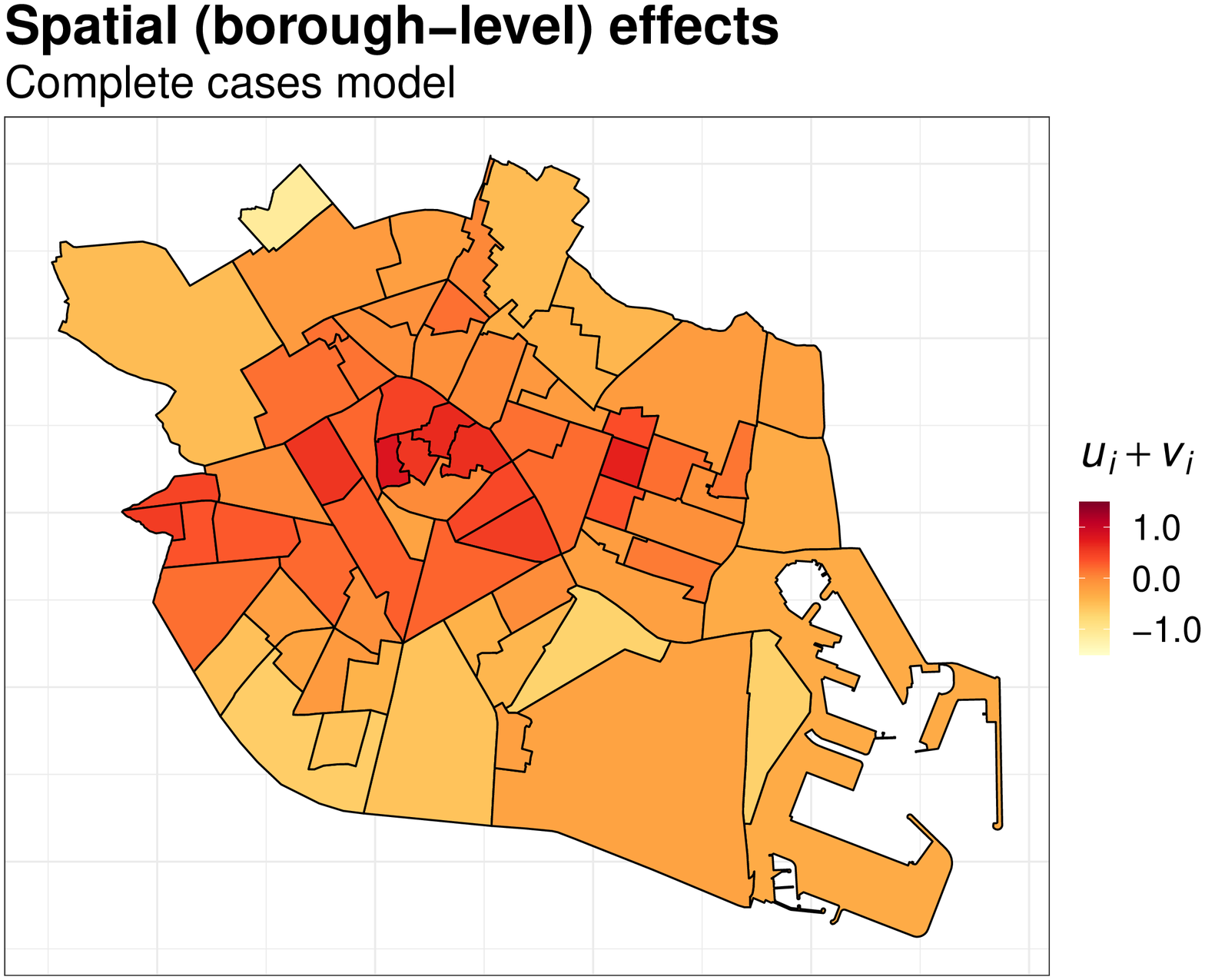}\label{fig:spatial_effects_impute_no_a}}
 \subfloat[]{\includegraphics[width=5cm,angle=-90]{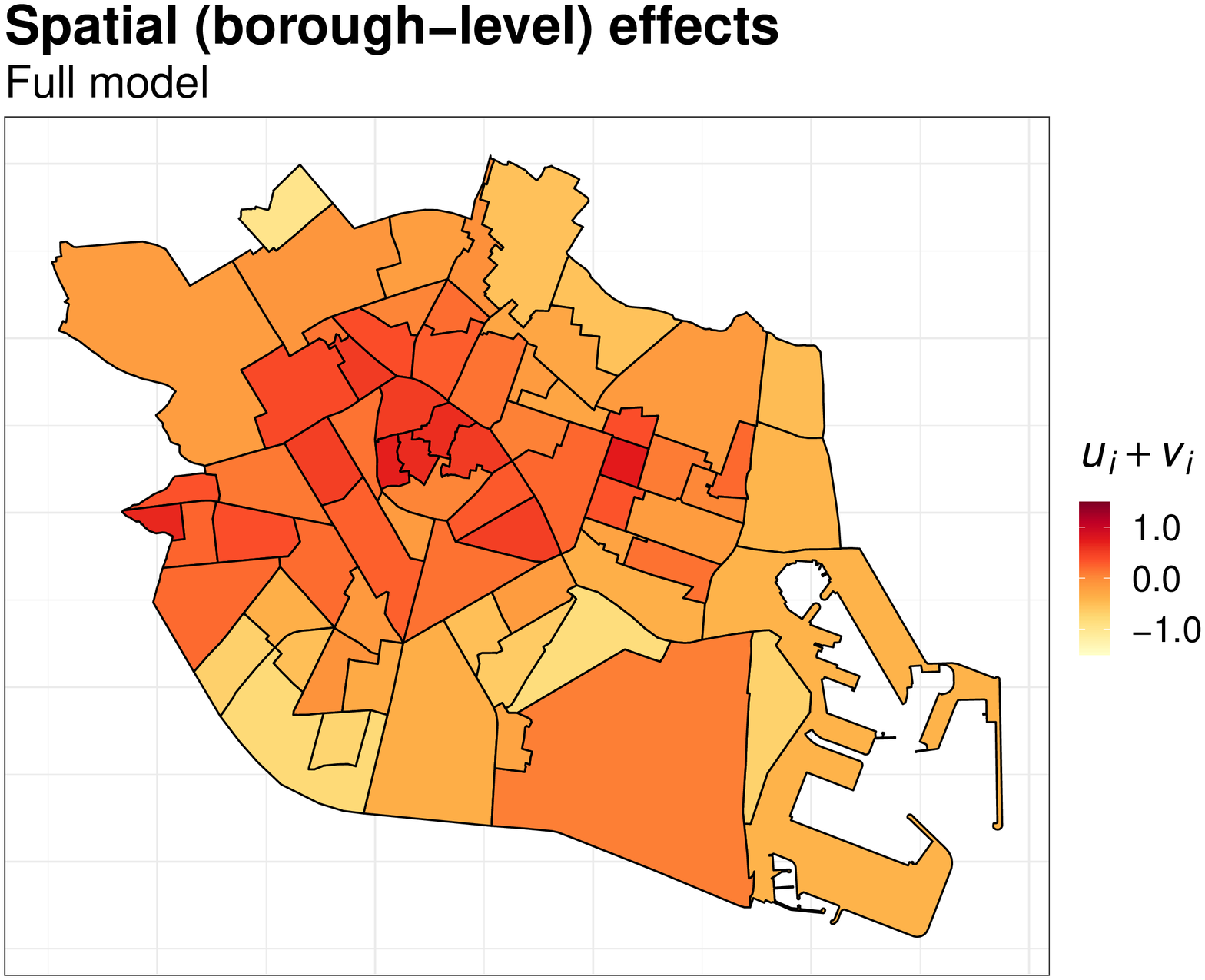}\label{fig:spatial_effects_impute_no_b}}
\caption{Spatial effect estimates (at the borough level) yielded by the complete cases model (a) and the full model (b)}
\label{fig:spatial_effects_impute_no}
\end{figure}

Finally, Figure \ref{fig:spatial_effects_impute_no} enables us to compare the spatial random effects estimates resulting from both models. In this case, the differences are minimal, being more noticeable in some neighbors located around the city center. In other words, these results suggest that the presence of temporally-uncertain events does not follow a spatial pattern for the dataset analyzed.

\subsubsection{Event time imputation}

\begin{figure}[htbp]
 \centering
 \includegraphics[width=8cm,angle=-90]{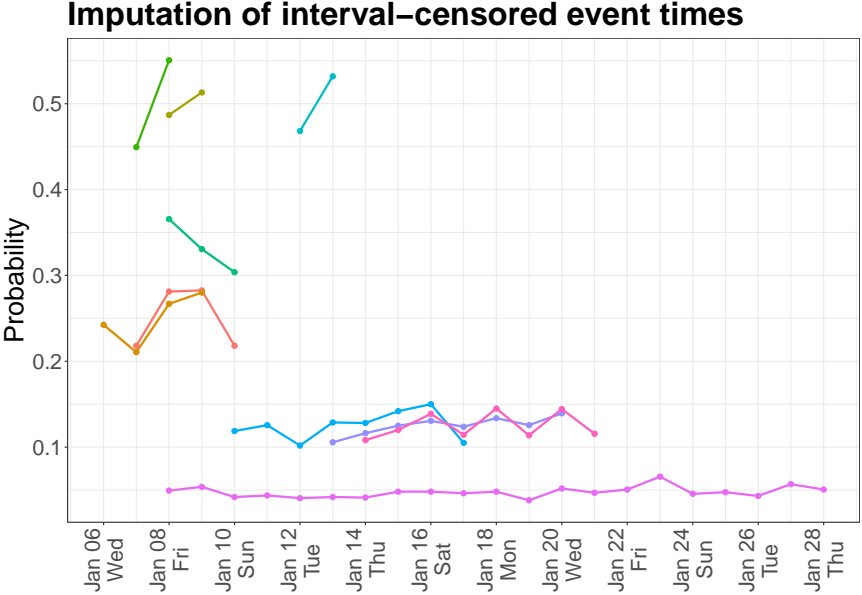}
\caption{Estimates of $p(t_{i}^{event}=t|D)$ for a selection of temporally-uncertain burglaries occurred from 6 January 2016 to 28 January 2016}
\label{fig:imputed_probabilities}
\end{figure}

The main advantage of the proposed full model is that it allows the inclusion of temporally-uncertain events in the analysis, following the aoristic approach. In this way, it avoids reducing the sample size (as occurs in the complete cases model) and prevents the potential error involved in imputing the time of the event. In addition, another advantage of the model is that it makes it possible to perform the imputation of event times, based on the posterior probability of each time unit (in our case, days) contained in the intervals that delimit the uncertainty existing for each event. Hence, Figure \ref{fig:imputed_probabilities} shows the values of $p(t_{i}^{event}=t|D)$ corresponding to a set of temporally-uncertain burglaries occurred from 6 January 2016 to 28 January 2016. In each case, $t$ varies from $t_{i}^{from}$ to $t_{i}^{to}$, therefore some probability of occurrence is assigned to each time unit contained in the interval. The values of $p(t_{i}^{event}=t|D)$ are based on all the information contained in the model, in terms of the fixed and random effects considered. In particular, the temporal uncertainty is connected to the day of the week effect and to the temporal random effect (the spatial random effect, on the other hand, is not influenced by the temporal uncertainty).  Thus, the values of $p(t_{i}^{event}=t|D)$ reflect the estimated crime risk for each day of the week and week within the study period. In particular, the values of $p(t_{i}^{event}=t|D)$ tend to be higher for the days of the week associated with higher risk, which are Friday and Saturday. This is clearly illustrated by some of the events shown in Figure \ref{fig:imputed_probabilities}, which exhibit a temporal uncertainty of 2, 3, or 4 days, including all or part of a weekend in the uncertainty window. For other events for which there is more uncertainty (the time window is wider), this effect is not as clear.

Therefore, the modeling approach described could be used as an imputation technique too, by simply considering the value $\arg \max_{t} p(t_{i}^{event}=t|D)$. It would be necessary to have a dataset in which the events have temporal uncertainty according to Police records and, at the same time, the exact temporal location has been provided by an external source, to assess the quality of this imputation method. 

\subsubsection{Model assessment}

\begin{figure}[htbp]
 \centering
 \subfloat[]{\includegraphics[width=5cm,angle=-90]{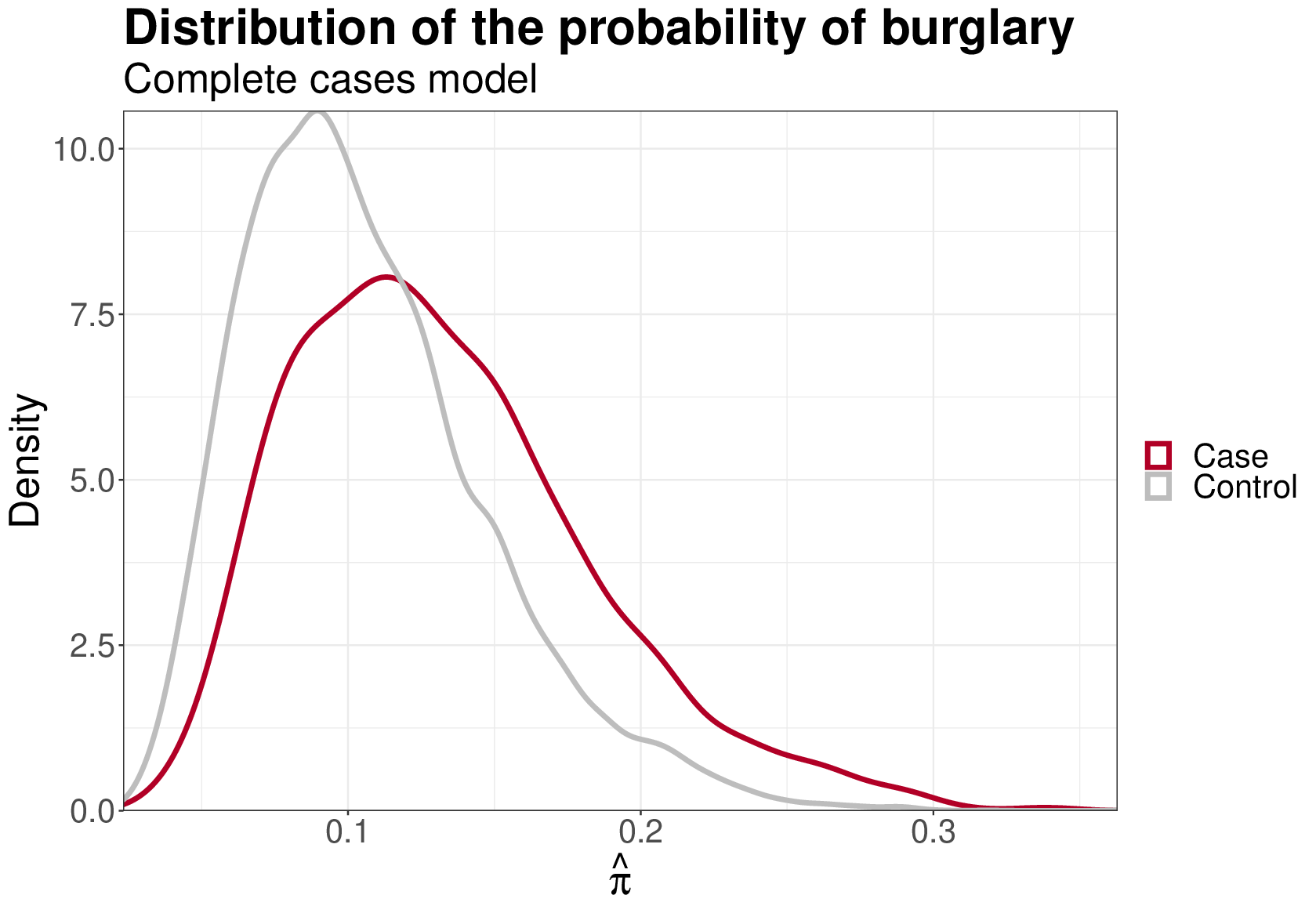}\label{fig:case_vs_control_probs_a}}
 \subfloat[]{\includegraphics[width=5cm,angle=-90]{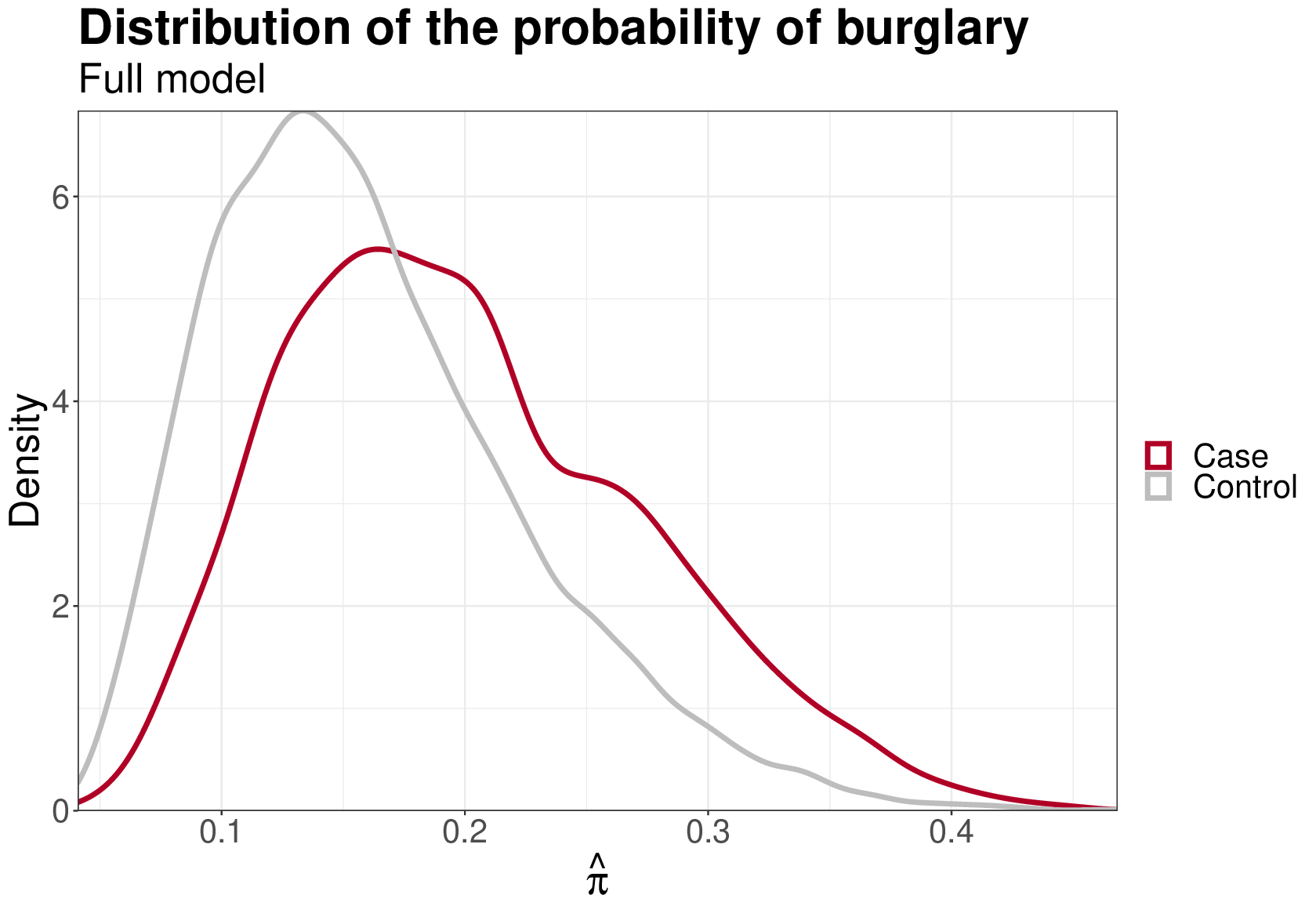}\label{fig:case_vs_control_probs_b}}
\caption{Distribution of the $\hat{\pi}_i$'s corresponding to cases (events) and controls for the complete cases model (a) and the full model (b)}
\label{fig:case_vs_control_probs}
\end{figure}

The first step to assess the quality of the models has been analyzing the distribution of the $\hat{\pi}_i$'s, computed as the mean of the posterior distribution $p(\pi_i|D)$. Figure \ref{fig:case_vs_control_probs} shows the distribution of the $\hat{\pi}_i$'s for both cases and controls, considering the complete cases (Figure \ref{fig:case_vs_control_probs_a}) and the full model (Figure \ref{fig:case_vs_control_probs_b}). It can be observed that both models can distinguish between cases and controls adequately, that is, the distribution of the $\hat{\pi}_i$'s presents higher values for the cases (the distribution is more displaced to the right). It can be also appreciated that the distribution of the $\hat{\pi}_i$'s covers larger values (for both cases and controls) in the case of the full model as a consequence of the greater proportion of cases in this model. This is something that has already been discussed when comparing the $\alpha$ parameters of the models and will be of importance again when evaluating the classification ability of the models, as will be shown later. 

\begin{figure}[htbp]
 \centering
 \includegraphics[width=8cm,angle=-90]{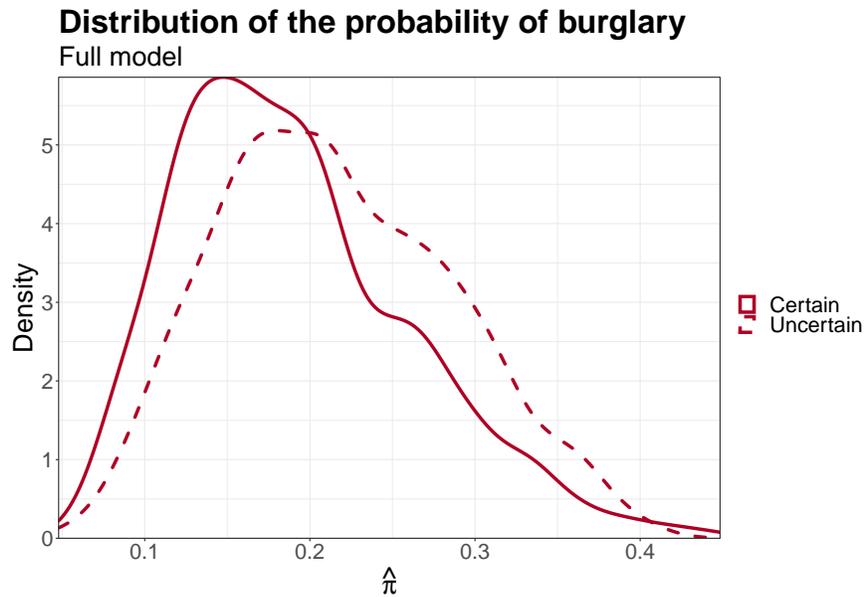}
\caption{Distribution of the $\hat{\pi}_i$'s corresponding to temporally-uncertain and certain (exact date known) events for the full model}
\label{fig:certain_vs_uncertain}
\end{figure}

Furthermore, Figure \ref{fig:certain_vs_uncertain} compares the distribution of the $\hat{\pi}_i$'s estimated through the full model for temporally-uncertain and certain (at the date level) events. The distribution of the $\hat{\pi}_i$'s corresponding to temporally-uncertain events presents a better behavior, in the sense that this distribution is more displaced towards larger estimates of $\pi_i$. This analysis allows us to verify that the temporally-uncertain observations have been adequately included in the model since they do not perform worse from the perspective of model fit than those that are not (in fact, they seem to perform better).

\begin{figure}[htbp]
 \centering
 \subfloat[]{\includegraphics[width=5cm,angle=-90]{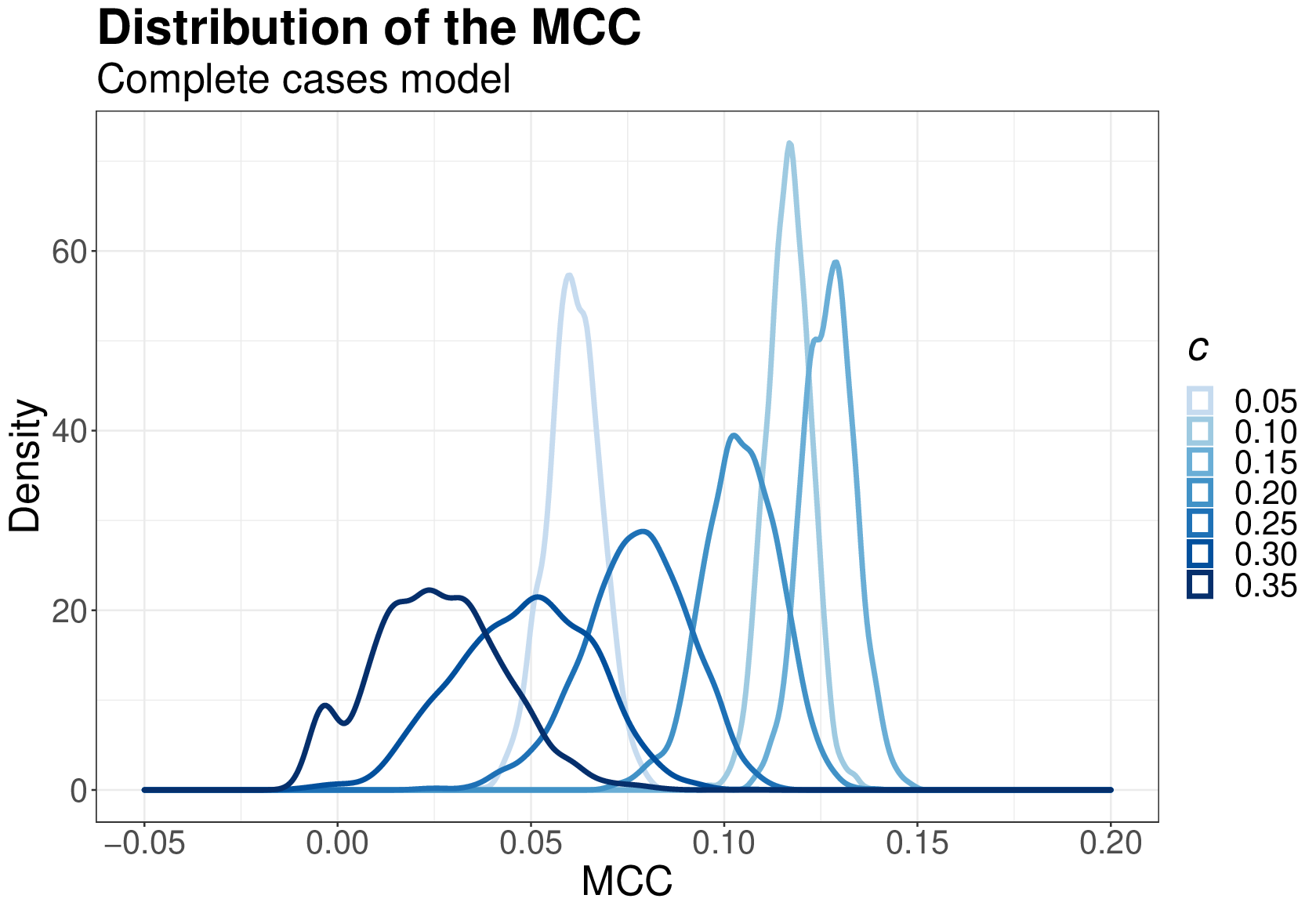}\label{fig:MCC_comparison_distribution_a}}
 \subfloat[]{\includegraphics[width=5cm,angle=-90]{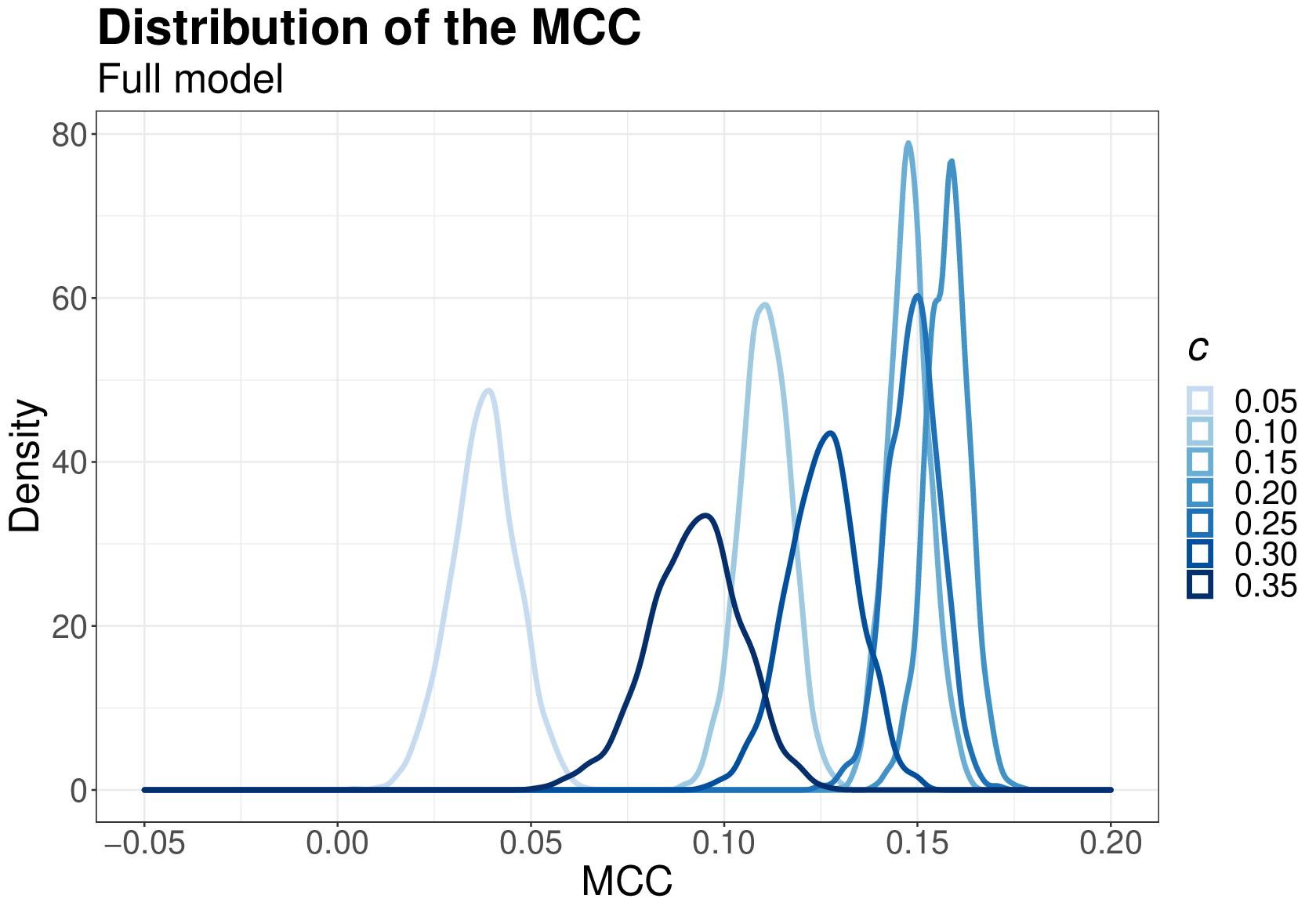}\label{fig:MCC_comparison_distribution_b}}
\caption{Distribution of the MCC derived from the sampled values of the posterior distribution $p(\pi_i|D)$, considering the complete cases model (a) and the full model (b). Several values of the cutoff probability, $c$, are tested and compared}
\label{fig:MCC_comparison_distribution}
\end{figure}

\begin{figure}[htbp]
 \centering
 \subfloat[]{\includegraphics[width=5cm,angle=-90]{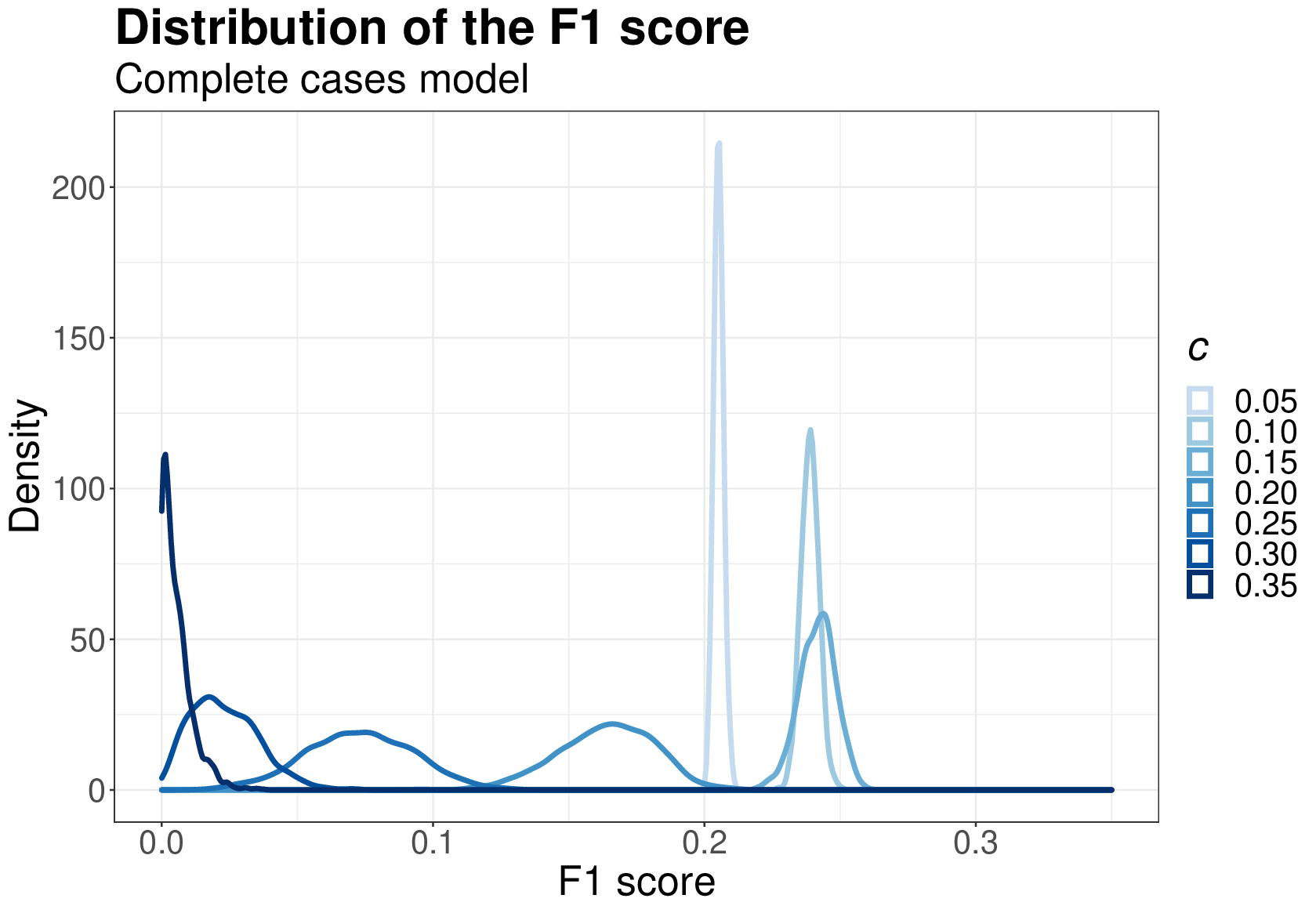}\label{fig:F1_comparison_distribution_a}}
 \subfloat[]{\includegraphics[width=5cm,angle=-90]{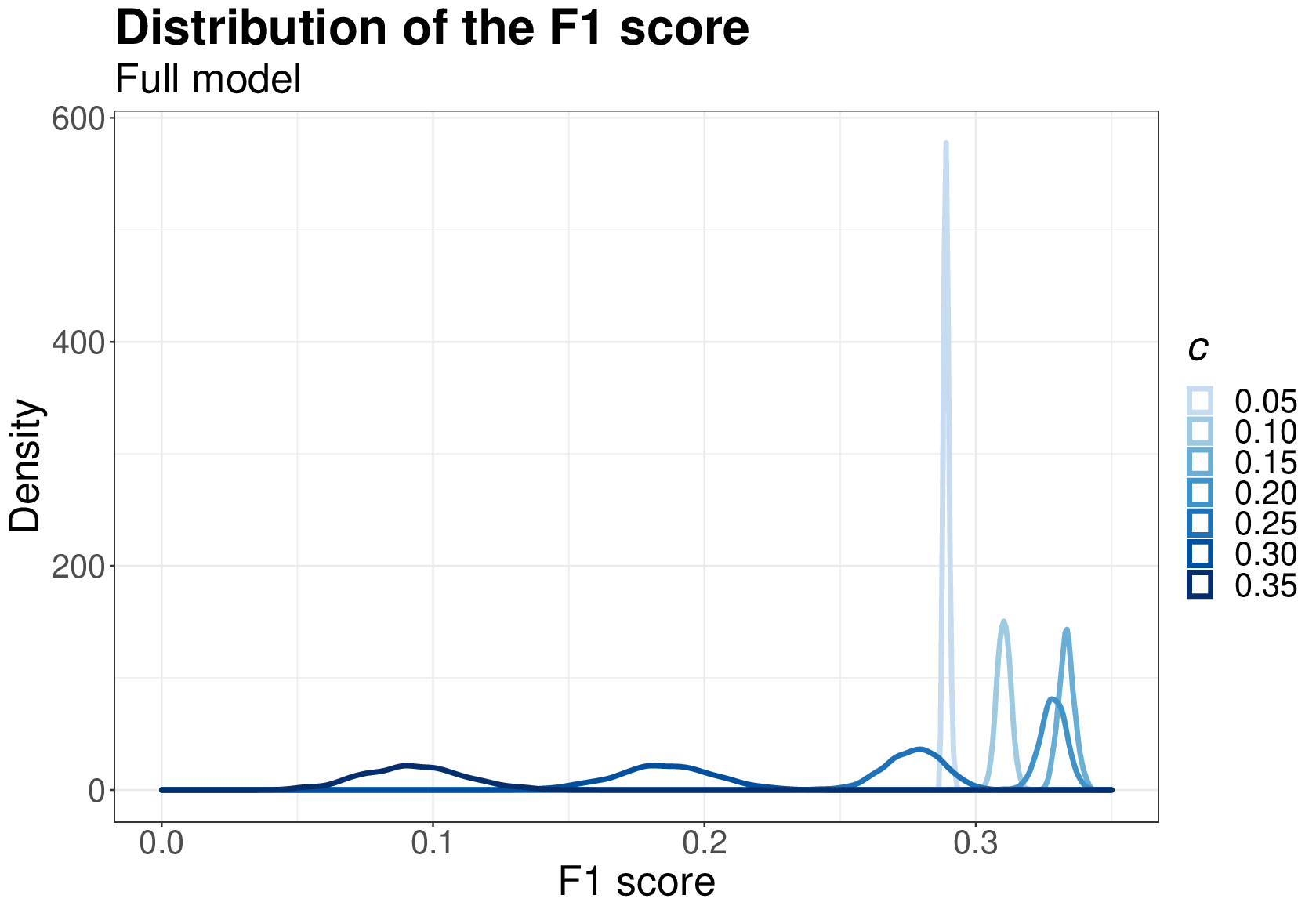}\label{fig:F1_comparison_distribution_b}}
\caption{Distribution of the F1 score derived from the sampled values of the posterior distribution $p(\pi_i|D)$, considering the complete cases model (a) and the full model (b). Several values of the cutoff probability, $c$, are tested and compared}
\label{fig:F1_comparison_distribution}
\end{figure}

Regarding the classification ability of the model, Figure \ref{fig:MCC_comparison_distribution} shows the distribution of the MCC, derived from the sampled values of the posterior distribution $p(\pi_i|D)$. The cutoff probability, $c$, is varied from 0.05 to 0.35, in steps of 0.05. Higher values of $c$ are discarded because the number of positive predictions becomes too low. This is a consequence of the fact that the dataset is imbalanced with few case observations (in comparison to the number of control observations), which causes model predictions to be less than 0.5. This is not an issue, we simply have to consider values of $c$ lower than 0.5 (which would be the common threshold for a balanced dataset). Figure \ref{fig:MCC_comparison_distribution} provides us with two conclusions of interest. First, the optimal values of $c$ are 0.15 and 0.20 for the complete cases and the full model, respectively (testing a finer partition of $c$ values would allow us to more accurately approximate the optimal value of $c$ in each case). The fact that the optimal value of $c$ is larger in the case of the full model is something that we might already expect since the proportion of cases is larger for the full dataset (this has been already discussed given the estimates of the $\alpha$ parameter for each of the models). Second, and more importantly, Figure \ref{fig:MCC_comparison_distribution} enables us to appreciate that MCC values tend to be higher in the case of the full model. Specifically, considering the optimal $c$ values, the MCC ranges from 0.114 to 0.140 (with 95\% credibility) in the case of the complete cases model (for $c=0.15$), whereas it ranges from 0.147 to 0.168 (with 95\% credibility) in the case of the full model (for $c=0.20$). For these choices of $c$, the resulting confusion matrices are $\left( \begin{smallmatrix} \mathrm{TP}&\mathrm{FP}\\ \mathrm{FN}&\mathrm{TN} \end{smallmatrix} \right)=\left( \begin{smallmatrix} 513&1903\\ 1086&11148 \end{smallmatrix} \right)$ and $\left( \begin{smallmatrix} \mathrm{TP}&\mathrm{FP}\\ \mathrm{FN}&\mathrm{TN} \end{smallmatrix} \right)=\left( \begin{smallmatrix} 1190&3151\\ 1434&9900 \end{smallmatrix} \right)$ for the complete cases and the full model, respectively. Similarly, the full model also performs better in terms of the F1 score, as shown in Figure \ref{fig:F1_comparison_distribution}. Specifically, the values of the F1 score are optimal for $c=0.15$, regardless of the model chosen. In the case of the complete cases model, the F1 score ranges from 0.227 to 0.254 (with 95\% credibility), while it ranges from 0.328 to 0.339 (with 95\% credibility) in the case of the full model.

\section{Discussion and conclusions}
\label{Discussion}

In this paper, a logistic regression model has been proposed for the analysis of crime data in the presence of temporally-uncertain observations, which are abundant for certain types of crimes. The aoristic method, which allows exploratory analysis of the data in this context, has been taken into account to incorporate such temporal uncertainty into the model. This is a natural approach considering the Bayesian treatment of missing data. The model implemented has allowed us to see how discarding temporally-uncertain observations in the analysis can lead to erroneous conclusions. Although this kind of modeling approach for dealing with interval-censored event observations has already been proposed in the literature \citep{reich2015partially}, this article has the novelty, to the best of the author's knowledge, of following the aoristic approach in a modeling context, while performing a complete comparison of the model proposed with the complete cases counterpart, which would be a typical choice.

There is still room for improvement for the model proposed. Indeed, the model could be enhanced by adding covariate information and interaction terms. For instance, a spatio-temporal interaction random effect or the interaction between the day of the week and the week within the year could be considered. The addition of such terms might lead to a more informative model, so the imputation of event times based on the posterior distribution of $t_{i}^{event}$ could be more realistic. As a drawback, increasing the complexity of the model by the inclusion of these terms could complicate model estimation, especially if the size of the available crime dataset is not very large. In addition, the way of handling the temporal uncertainty of the events could be extended to other types of models in which the spatial effects are not based on an areal structure, such as point process models \citep{mohler2011self,shirota2017space}.

Another important aspect is that when discussing the results provided by the complete cases and the full model, it has been implicitly assumed that the aoristic analysis of the data gives a true picture of the temporal distribution of the data. However, as pointed out by \cite{mulder2019bayesian}, the aoristic analysis might tend to overdisperse the temporal distribution of the events. Thus, by assigning the same weight to each temporal unit within the observation window, we might be assuming too much uncertainty (variability), even though it results in a natural approach if we have no prior knowledge about the true temporal location of the event. The proposed model allows the inclusion of prior information about the events, in the standard way used in Bayesian inference. For instance, a specific prior distribution could be assigned to those events for which there might be some intuition (by the Police or the property owners themselves) about their actual temporal location, or some non-uniform distribution that might be closer to reality could be tested. In fact, future studies could make use of a truncated Normal distribution with the mean located at the midpoint location, or even at a location closer to the start (or end) of the interval. This could reveal, in some cases, whether events tend to cluster temporally in the initial part of the time window, which could be explained in case the burglars have been watching the owners and took advantage of their departure from the home. 


Finally, to better assess the potential of the full model for predicting the true temporal location of the temporally-uncertain events, a dataset including both the interval-censored temporal locations recorded by the Police (according to the information provided by the owners or other residents) and the actual (exact) temporal locations derived from other sources would be required. For instance, in the case study conducted by \cite{ashby2013comparison}, closed-circuit television camera images were used to determine the exact temporal locations of the events under study. Unfortunately, this kind of dataset is really scarce. 







\clearpage

\bibliography{Bibliography}

\end{document}